\documentclass[apj]{emulateapj}

\def\kms{\hbox{$~$km$~$s$^{-1}$}}
\def\l{\ifmmode\lambda\else$\lambda$\fi}
\def\snia{SN~Ia}
\def\sneia{SNe~Ia}
\def\caone{Ca\,{\sc I}}
\def\catwo{Ca\,{\sc II}}
\def\fetwo{Fe\,{\sc II}}
\def\mgtwo{Mg\,{\sc II}}
\def\naone{Na\,{\sc I}}
\def\naoned{Na\,{\sc I}\,D}
\def\naonedone{Na\,{\sc I}\,D$_1$}
\def\naonedtwo{Na\,{\sc I}\,D$_2$}
\def\sitwo{Si\,{\sc II}}

\slugcomment{Accepted for publication in Ap.~J.}
\shorttitle{Variable \naoned\ in \sneia}
\shortauthors{Blondin et al.}

\begin{document}

\title{A Second Case of Variable N\lowercase{a}~I~D Lines in a
  Highly-Reddened Type I\lowercase{a} Supernova} 

\author{
S.~Blondin,\altaffilmark{1,2}
J.~L.~Prieto,\altaffilmark{3}
F.~Patat,\altaffilmark{2} 
P.~Challis,\altaffilmark{1} 
M.~Hicken,\altaffilmark{1} 
R.~P.~Kirshner,\altaffilmark{1} 
T.~Matheson,\altaffilmark{4}
M.~Modjaz\altaffilmark{5} 
}


\altaffiltext{1}{Harvard-Smithsonian Center for
  Astrophysics, 60 Garden Street, Cambridge, MA 02138;
  {sblondin,pchallis,mhicken,rkirshner@cfa.harvard.edu} .}
\altaffiltext{2}{European Southern Observatory,
  Karl-Schwarzschild-Strasse 2, D-85748 Garching, Germany;
  {sblondin,fpatat@eso.org} .}
\altaffiltext{3}{Dept. of  Astronomy, The Ohio State
  University, 140 W. 18th Ave., Columbus, OH 43210;
  {prieto@astronomy.ohio-state.edu} .}
\altaffiltext{4}{National Optical Astronomy
  Observatory, 950 North Cherry Avenue, Tucson, AZ 85719-4933;
  {matheson@noao.edu} .} 
\altaffiltext{5}{Department of Astronomy, University of
  California, Berkeley, CA 94720-3411; {mmodjaz@astro.berkeley.edu} .}


\begin{abstract}
Recent high-resolution spectra of the Type Ia SN~2006X have revealed
the presence of time-variable and blueshifted \naoned\ features,
interpreted by Patat et al. as originating in circumstellar material 
within the progenitor system. The variation seen in SN~2006X induces
relatively large changes in the total \naoned\ equivalent width
($\Delta {\rm EW} \sim 0.5$\,\AA\ in just over two weeks), that would
be detectable at lower resolutions. We have used a large data set 
comprising 2400 low-resolution spectra of 450 Type Ia supernovae
(\sneia) obtained by the CfA Supernova Program to
search for variable \naoned\ features. Out of the 31
\sneia\ (including SN~2006X) in which we could have detected similar
EW variations, only one other (SN~1999cl) shows variable \naoned\
features, with an even larger change over a similar $\sim10$-day
timescale ($\Delta {\rm EW} = 1.66\pm0.21$\,\AA). 
Interestingly, both 
SN~1999cl and SN~2006X are the two most highly-reddened objects in our
sample, raising the possibility that the variability is connected to 
dusty environments.
\end{abstract}



\section{Introduction}\label{sect:intro}

Type Ia supernovae (\sneia) are thought to result from the
thermonuclear runaway in a carbon-oxygen white dwarf (WD) star
\citep{Hoyle/Fowler:1960}, having grown in mass to the Chandrasekhar
critical mass \citep{Chandrasekhar:1931}. How the WD actually approaches
this critical mass is still debated
\citep[e.g.,][]{Parthasarathy/etal:2007,Tutukov/Fedorova:2007}, 
but viable hypotheses include the accretion of material from a
non-degenerate binary companion (the ``single-degenerate'' scenario;
\citealt{Whelan/Iben:1973}), or a merger process involving another
white dwarf star (the ``double-degenerate'' scenario;
\citealt{Iben/Tutukov:1984,Webbink:1984}).

While there is no direct empirical evidence for either scenario, 
observations suggest the existence of multiple progenitor
channels. For instance, the distribution of \snia\ peak luminosities is
different in early-type and late-type galaxies (the most luminous \sneia\
are found only in late-type galaxies; e.g., 
\citealt{Hamuy/etal:2000,MLCS2k2}).
Several studies have modeled the \snia\ rate with two
components: one proportional to the star formation rate, the other
proportional to the total stellar mass 
\citep{Mannucci/etal:2005,Scannapieco/Bildsten:2005,Sullivan/etal:2006}.
The dependence on the star formation rate is evidence that at least
some \sneia\ originate in young stellar populations, and confirms
earlier claims of an enhanced \snia\ rate in star-forming
galaxies
\citep[e.g.,][]{Oemler/Tinsley:1979,vandenBergh:1990,Cappellaro/etal:1997}.
More recently, \cite{Pritchet/Howell/Sullivan:2008} find that the
\snia\ rate is $\sim1$\% of the WD formation rate, independent of
star-formation history, and conclude that the single-degenerate
channel alone is not a viable model for all \snia\ progenitor systems
(see also \citealt{Yungelson/Livio:2000,Greggio:2005}).

The double-degenerate scenario has been invoked as a plausible
interpretation for a few unusually luminous \sneia\ (e.g., SNLS-03D3bb,
\citealt{Howell/etal:2006}; SN~2006gz, \citealt{Hicken/etal:2007}).
However, several hydrodynamical models suggest that double-WD mergers
cause an off-center carbon ignition that leads to the conversion of
C+O into O+Ne+Mg (e.g., \citealt{Saio/Nomoto:2004}), and
subsequently to an accretion-induced collapse to a neutron star
(e.g., \citealt{Bravo/Garcia-Senz:1999}; although see
\citealt{Yoon/Podsiadlowski/Rosswog:2007}). 
Radiation-hydrodynamics simulations of such O+Ne+Mg cores predict a
weak $\lesssim$10$^{50}$\,erg) explosion if rotation is absent, powered
primarily by neutrino-energy deposition
\citep{Kitaura/etal:2006,Dessart/etal:2006}. However, angular momentum
is expected to be abundant in such accretion-induced collapses,
resulting in a strong aspherical explosion. The predicted low
$^{56}$Ni yields and ejecta mass would make these events optically
dim, quite unlike \snia\ explosions \citep{Dessart/etal:2007}.

A direct confirmation of the single-degenerate scenario would be the
detection of circumstellar material (CSM) arising from the transfer of 
matter to the WD by its non-degenerate binary companion. The
interaction of the \snia\ ejecta with the CSM would probably lead
to detectable radio and X-ray emission, but observations at these
wavelengths yield only upper limits
\citep{Panagia/etal:2006,Hughes/etal:2007}. Nonetheless, strong
hydrogen emission has been detected in two \sneia\ so far (SN~2002ic,
\citealt{Hamuy/etal:2003}; SN~2005gj,
\citealt{Aldering/etal:2006,Prieto/etal:2007}; although see
\citealt{Benetti/etal:2006} and \citealt{Trundle/etal:2008} for a
different interpretation), interpreted as
evidence for ejecta-CSM interaction, and hence for a single-degenerate
scenario (although see \citealt{Livio/Riess:2003}). However, the lack
of similar detections in other \sneia\ suggests that these are rare
events.

Moreover, models
suggest that a significant amount of material is stripped from the
companion's envelope during the explosion
\citep{Chugai:1986,Livne/Tuchman/Wheeler:1992,Marietta/Burrows/Fryxell:2000,Meng/Chen/Han:2007},
which should lead to an observable signature in the form of a narrow
H$\alpha$ emission line in the nebular spectra. No such signature has 
yet been detected \citep{Mattila/etal:2005,Leonard:2007}. A recent
hydrodynamical investigation of the interaction between the ejecta of
a Type Ia supernova and a main sequence binary companion
\citep{Pakmor/etal:2008} finds that the theoretical predictions for
the mass of stripped hydrogen are consistent with the observational
limits.

In a recent study, \cite{Patat/etal:2007a} reported on the possible
detection of circumstellar material in the normal Type Ia SN~2006X.
Using high-resolution spectra at four epochs out to $\sim2$ months
past explosion, they observed a complex evolution in blueshifted
absorption features associated with the
\naoned\ doublet. Because of the lack of similar evolution in the
\catwo\,H\&K doublet, they concluded that the time evolution of the
\naoned\ complex is not due to line-of-sight effects in interstellar gas.
Instead, the variation was attributed to
ionization changes in circumstellar material (CSM) within the
progenitor system, due to radiation from the supernova. The
expansion velocities of this CSM ($\sim50$\,\kms) were compatible with
wind velocities for early red-giant stars, thereby favoring a
single-degenerate progenitor scenario for SN~2006X. 

More recent studies of SN~2000cx and SN~2007af using similar multiepoch
high-resolution data have not detected any variable \naoned\ features. 
No \naoned\ absorption is detected in
SN~2000cx to the level of a few m\AA\
\citep{Patat/etal:2007b}, setting an upper limit of
$2\times10^{10}$\,cm$^{-2}$ on the column density of the absorbing
material. In SN~2007af, \cite{Simon/etal:2007} place an upper limit of
$\sim10$\,m\AA\ on the EW of circumstellar \naoned\ lines,
corresponding to a column density of $6\times10^{10}$\,cm$^{-2}$.

It is therefore still unclear whether the variability in the \naoned\
line is a common feature amongst \sneia, or if SN~2006X is a unique
case. \cite{Chugai:2008a} argues that the structure observed by
\cite{Patat/etal:2007a} cannot originate in circumstellar material,
based on the expected low optical depth of \naoned\ ($\tau < 10^{-3}$)
in the spherically-symmetric wind of a typical red giant (RG) star. 

With a large sample of multiepoch high-resolution spectra of \sneia,
one could address these questions directly. However, such data require
dedicated programs on 8--10\,m telescopes, and signal-to-noise ratio (S/N)
constraints restrict this study to bright events ($m_V \lesssim 15$ at
peak).

In this paper, we instead use a large database of low-resolution
spectra obtained through the CfA Supernova Program to
investigate changes in the \naoned\ doublet with time. The variations
in the \naoned\ profile in SN~2006X induced changes in its equivalent
width (EW) of $>0.5$\,\AA\ between $-2$ and +14\,d from maximum
light. We have a unique set of hundreds of \sneia\ whose spectra we
can examine for EW variations like those seen in SN~2006X. 
The statistical power of such a large sample enables us to place
meaningful constraints on the frequency (and nature) of variable
\naoned\ features in \sneia.

We show that SN~2006X is not the only supernova with variable
\naoned\ absorption, and the nature of the variable absorption we
detect leads us to think that the explanation given by
\cite{Patat/etal:2007a} for SN~2006X may not apply in all
cases.  What's more, the absence of variable absorption in most of the
\sneia\ for which we have adequate data is an important clue to the
nature of SN Ia progenitors.

The paper is organized as follows. In \S~\ref{sect:data} we
present the spectroscopic sample used in this study. We use a Monte
Carlo simulation to estimate the systematic error in our EW
measurements in \S~\ref{sect:mc}. The results are presented in
\S~\ref{sect:res} and discussed in \S~\ref{sect:disc}. Conclusions
follow in \S~\ref{sect:ccl}.


\section{Spectroscopic Data}\label{sect:data}

To investigate the variability of \naoned\ features in \sneia\ we have
used a large spectroscopic data set obtained through the CfA Supernova
Program (P.I.: R.~Kirshner). Since 1994, we have obtained $\sim 2400$
optical spectra of $\sim 450$ low-redshift ($z \lesssim 0.05$) \sneia\
with the 1.5\,m Tillinghast telescope at FLWO using the FAST
spectrograph \citep{Fabricant/etal:1998}. The setup used (300-line/mm
grating and 3\arcsec\ slit) yields a
typical FWHM resolution of $\sim6$\,\AA\ ($\sim300$\,\kms\ at
\naoned). Several spectra were 
published in studies of specific supernovae (e.g., SN~1998bu; 
\citealt{Jha/etal:1999}), while 432 spectra of 32 \sneia\ have
recently been published by \cite{Matheson/etal:2008}. We also have
complementary multi-band optical photometry for most objects
\citep{Riess/etal:1999a,Jha/etal:2006,Hicken/etal:2009a}. All
published data are available via the CfA Supernova 
Archive\footnote{http://www.cfa.harvard.edu/supernova/SNarchive.html}.

In SN~2006X, the bulk of the variation in the \naoned\ EW occurs
before $\sim15$~d past maximum light, with a timescale of $\sim10$~d
\citep{Patat/etal:2007a}. In order to 
investigate the occurrence of similar cases in our data, we only
consider objects for which we have at least one spectrum before +10~d
past maximum light, and for which the spectroscopic coverage spans at
least 10~d. The resulting sample comprises 1826 spectra of 168 \sneia.

All the spectra were obtained with the same telescope and instrument,
and reduced in a consistent manner (see \citealt{Matheson/etal:2008}
for details). The uniformity of this data set is unique and
enables an accurate estimate of systematic errors in our
measurements.


\section{Monte Carlo Error Estimation}\label{sect:mc}

Detecting small \naoned\ equivalent-width variations is a challenging
task with low-resolution spectra. At the resolution of the FAST
spectrograph (${\rm FWHM} \approx 6$\,\AA), both D$_1$ (5895.92\,\AA)
and D$_2$ (5889.95\,\AA) lines comprising the \naoned\ doublet are
blended together. Any variation in the strength of (weaker)
blueshifted components (as observed in SN~2006X;
\citealt{Patat/etal:2007a}) will then result in a small fractional
change in the total (D$_1$+D$_2$) EW of a single \naoned\ absorption
feature.

In what follows we use a Monte Carlo simulation to determine the
minimum detectable EW variation as a function of resolution and
signal-to-noise ratio (S/N). We use the results of this simulation to
estimate the number of objects for which we would detect EW
variations comparable to those seen in SN~2006X.

\subsection{Simulation presentation}

We generate synthetic \naoned\ profiles consisting of one main doublet
and several blueshifted sub-components. In the case of SN~2006X these
were interpreted as being of interstellar and circumstellar origin,
respectively \citep{Patat/etal:2007a}. The simulation parameters are
summarized in Table~\ref{tab:mcparams} and explained below.

\begin{deluxetable}{lc}
\tablewidth{0pt}
\tablecaption{\label{tab:mcparams}Simulation parameters}
\tablehead{
\colhead{Parameter} & \colhead{Range}}
\startdata
Doppler parameter (\kms) & $b=10$ \\ 
Optical thickness of main \naonedtwo\ line\tablenotemark{a} & $-1 \le \log{\tau_0} \le 5$ \\
Number of \naoned\ sub-components & $1 \le n_{\rm sub} \le 5$ \\
Optical thickness of sub-components\tablenotemark{a} & $-2 \le \log{\tau_{0,\rm{sub}}} \le 0$ \\
Blueshifts of sub-components (\kms) & $0 \le v_{\rm sub} \le 100$ \\
SN age (days from $B$-band maximum) & $-15 \le t_{\rm SN} \le +85$ \\
Resolution\tablenotemark{b} (\AA) & $0.1 \le {\rm FWHM} \le 15.0$ \\
Signal-to-noise ratio (per pixel) & $1 \le {\rm S/N} \le 200$
\enddata
\tablenotetext{a}{$\tau_0$ is the optical thickness at line center.}
\tablenotetext{b}{We assume there are 4.2 pixels per FWHM.}
\end{deluxetable}

We start by generating the main \naoned\ doublet at high resolution
(${\rm FWHM} = 0.1$\,\AA), assuming pure absorption and a Voigt
profile shape. The width of the profile is set by the so-called
Doppler parameter $b$, which is a measure of the thermal and turbulent
motions within interstellar clouds. In all that follows we set
$b=10$\,\kms\ (this choice has negligible impact on our results). We
parametrize the doublet according to the optical thickness of the
stronger D$_2$ line, $\tau_0$. (The optical thickness of the
D$_1$ line is simply $\tau_0/2$ by virtue of the ratio of oscillator
strengths).

We then generate a random number of up to five sub-components in a
similar fashion, blueshifted by up to 100\,\kms\ with respect to the
main component. This corresponds to the velocity range of winds of
early red giant or late subgiant stars \citep[e.g.,][]{Judge/Stencel:1991},
that was invoked by \cite{Patat/etal:2007a} to explain the observed
blueshifts of the various \naoned\ sub-components observed in
SN~2006X. In principle, these components could form in faster
winds (e.g., from main sequence or compact helium stars), or through
the interaction of remnant shells of successive novae with the wind of
the donor star, in which case the blueshifts could be up to an order
of magnitude larger. However, we deliberately limit our investigation
to match the only observed case of SN~2006X.

\begin{figure}
\plotone{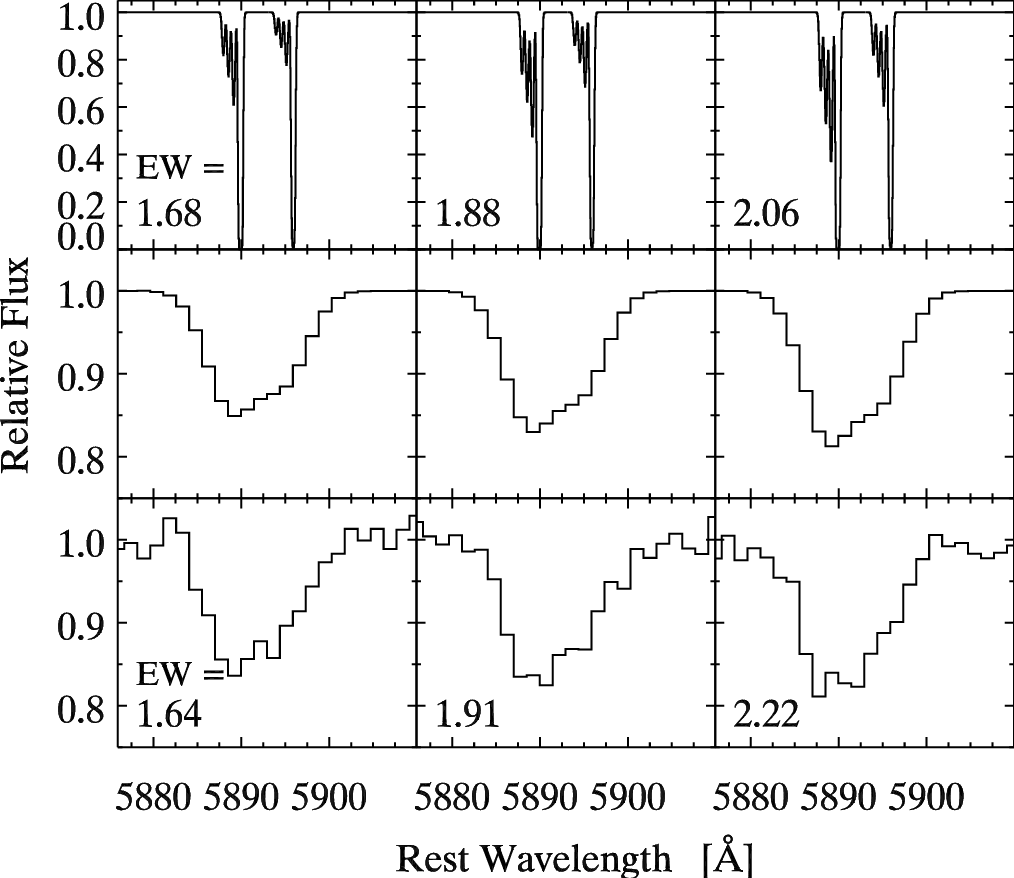}
\caption{\label{fig:mcfig}
{\it Top:} Example input normalized line profiles, consisting of the
main \naoned\ doublet and three weaker blueshifted sub-components. The
strength of these components increases from left to right. The number
in the lower-left corner is the equivalent width (in \AA) of the
entire profile.
{\it Middle:} As above but at the resolution of our spectroscopic data
set (FWHM\,$\approx6$\,\AA). Note the different ordinate range.
{\it Bottom:} As above with added Poisson noise (S/N\,$=60$ per pixel).
We also report the equivalent width of the degraded
profile (with a typical error $\sim 0.1$\AA; see
\S~\ref{sect:simres}).
}
\end{figure}

The top panels of Fig.~\ref{fig:mcfig} shows example high-resolution
synthetic line profiles. In all cases the stronger main \naoned\
components are identical (with $\log{\tau_0}=1$), while the strength
of the three (weaker) blueshifted components increases from left to
right, resulting in an increase of the total EW from $\sim1.7$ to
$\sim2.1$\,\AA. Note that, in our simulation, $\tau_0$ spans six decades
(Table~\ref{tab:mcparams}) , such that we also investigate cases where
the main \naoned\ component is weaker than the blueshifted
sub-components. We further consider cases where the sub-components
are at the same wavelength as the main component (``blueshift'' of
0\,\kms), such that the main \naoned\ lines vary while the blueshifted
sub-components do not. This situation is particularly relevant to an
example of variable \naoned\ we present in \S~\ref{sect:varna}.

\epsscale{1.1}
\epsscale{1.0}

In measuring equivalent widths the definition of the underlying
supernova pseudo-continuum is a potential source of systematic
error. We account for this by multiplying the normalized \naoned\
profiles by SN template spectra (from \citealt{Hsiao/etal:2007}) at
ages between $-15$ and +85~d from maximum light. The age distribution
is matched to that of our data set. Over this time interval the
underlying pseudo-continuum at the \naoned\ position changes
significantly. In \S~\ref{sect:simres} we show that this has
negligible impact on the measured equivalent width.

The result is then degraded in resolution through convolution with a
Gaussian profile, and resampled to 4.2 pixels per FWHM resolution to
match our data. We allow for resolutions as low as 15\,\AA, but
illustrate the typical resolution of our spectroscopic data set
(FWHM\,$\approx6$\,\AA) in the middle panels of
Fig.~\ref{fig:mcfig}. While we are unable to fully resolve individual
components of the \naoned\ doublet, the overall increase in the total
EW is clearly visible.

Last, we degrade the S/N by adding Poisson noise to the lower
resolution spectrum. Again, the overall increase in the total \naoned\
EW is visible in the bottom panels of Fig.~\ref{fig:mcfig}.

We then measure the equivalent width as follows: we define continuum
regions on either side of the D$_2$ and D$_1$ lines (or of the single
\naoned\ line when the doublet is unresolved) spanning a maximum of
20\,\AA. We assume the line to span three times the width of the
Doppler (Gaussian) core, or 
$3\times{\rm FWHM}/2\sqrt{2\ln{2}}\approx1.3\times{\rm FWHM}$, plus
an additional $\sim2$\,\AA\ on the blue side to account for the
possible presence of blueshifted sub-components. The continuum is then
determined by fitting a second-order polynomial to the continuum
regions, and the equivalent width is computed over the line span
defined in the usual way.

\subsection{Simulation results}\label{sect:simres}

The total equivalent width measured on the synthetic profile degraded in
resolution and S/N is compared with the known input total EW. We then
investigate the variation of the mean and rms of the difference as a
function of S/N. 

\begin{figure}
\plotone{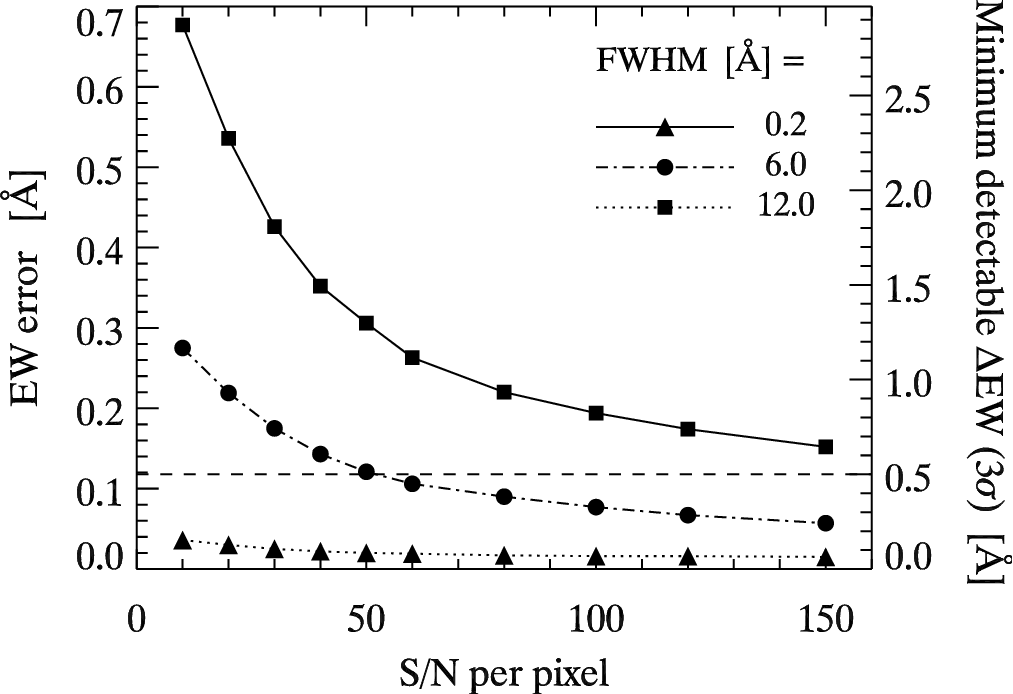}
\caption{\label{fig:ewerr}
EW measurement error as a function of S/N for three FWHM
resolutions. The dash-dotted curve corresponds to the 
resolution of our data set (${\rm FWHM}\approx 6$\,\AA). The ordinate
axis on the right indicates the corresponding 3$\sigma$ error on the
difference between EW measurements. The horizontal dashed line
corresponds to a minimum detectable EW difference of 0.5\,\AA.
}
\end{figure}

We summarize the simulation results in Fig.~\ref{fig:ewerr} for the
case where $\log{\tau_0} = 1$. We show the rms residuals (or
measurement error) of the difference between the true and measured
total \naoned\ EW for three FWHM resolutions (0.2, 6.0, and
12.0\,\AA) as a function of S/N. As expected, the error decreases with
increasing S/N and resolution. For a $3\sigma$ detection of an EW
variation of $\gtrsim0.5$\,\AA, as seen in SN~2006X between $-2$ and
+14\,d from maximum light \citep{Patat/etal:2007a}, the measurement
error must be $\lesssim 0.1$\,\AA, which in turn implies ${\rm S/N}
\gtrsim 50$ per pixel at a FWHM resolution of 6\,\AA. Applying this
S/N cut to our data further reduces the sample presented in
\S~\ref{sect:data} to 294 spectra of 31 \sneia.

\epsscale{1.1}
\epsscale{1.0}

We find that the strength of the main \naoned\ component has
a small systematic effect on our measurement accuracy. The mean
difference between the input and measured EW is $\lesssim0.04$\,\AA\
for $\log{\tau_0} \lesssim 4$ and ${\rm S/N} > 10$ per pixel 
(at a resolution of ${\rm FWHM}= 6$\,\AA),
independent of the number and strength of the blueshifted \naoned\
sub-components. 
The shape of the underlying pseudo-continuum at \naoned\
(which changes significantly with the age of the supernova)
contributes no additional error. The
measurement error due to resolution alone (i.e. the rms residuals for
${\rm FWHM} = 6$\,\AA\ at infinite S/N) is $\sim0.01$\,\AA. In what
follows, we assume a total systematic error $\sigma_{\rm sys} =
0.05$\,\AA.


\section{Results}\label{sect:res}

Our final sample comprises \sneia\ that span a large range of known
light-curve properties. In particular, the light-curve decline rate
parameter $\Delta m_{15} (B)$ \citep{Phillips:1993} ranges from
0.83 for the overluminous SN~1995al \citep{Phillips/etal:1999} to
1.48 for the slightly underluminous SN~2005am (Folatelli \& Hamuy
2008, private communication; see also
\citealt{Li/etal:2006}). However, none of the \sneia\ in our sample
has spectra or light curves similar to the more underluminous
SN~1991bg ($\Delta m_{15} (B) = 1.93$; \citealt{Phillips/etal:1999}).
 
We have measured the total \naoned\ EW on all 294 spectra using the
same technique as for our simulated data. For each measurement, we use
the corresponding error spectrum to derive the associated statistical
error, which is added in quadrature to the systematic error
derived from our simulation ($\sigma_{\rm sys} = 0.05$\,\AA; see
\S~\ref{sect:simres}).

We take special care to include only the host-galaxy \naoned\ line
when measuring the total EW. For redshifts $cz \lesssim 800$\,\kms,
the Galactic \naoned\ line contributes to the host-galaxy line region
defined in \S~\ref{sect:mc}. This affects only one \snia\ in our final
sample, SN~1994D, whose host galaxy NGC 4256 is at $cz = +592$\,\kms\
\citep{UZC}. However, \cite{King/etal:1995} report a recession
velocity of +880\,\kms\ {\it at the SN position}, which is the value we
assume in this study. Nonetheless, the presence of Galactic
high-velocity clouds inferred from high-resolution spectra (up to
+250\,\kms; \citealt{Ho/Filippenko:1995,King/etal:1995}) causes some
overlap in our spectra with the \naoned\ absorption in NGC 4256 in the
SN line of sight, and our EW measurements for SN~1994D are
systematically biased towards higher values (${\rm
  EW}=0.32_{-0.06}^{+0.34}$\,\AA, cf. $0.087\pm0.002$\,\AA\ reported
by \citealt{Ho/Filippenko:1995} for the NGC 4256 component; see
Table~\ref{tab:ewebv}). All other 30 \sneia\ are at $cz > 800$\,\kms,
and we interpolate over any Galactic \naoned\ for the EW determination.

\subsection{Variable \naoned\ in the highly-reddened SN~1999cl}\label{sect:varna}

The single example of variable \naoned\ lines observed to
date, SN~2006X, is too slender a foundation on which to construct a
general model for the variation. Nevertheless, we can test the null
hypothesis of a constant EW against our data for each supernova.
Rejecting this hypothesis constitutes detecting variability for an
individual supernova.

\begin{figure*}   
\plotone{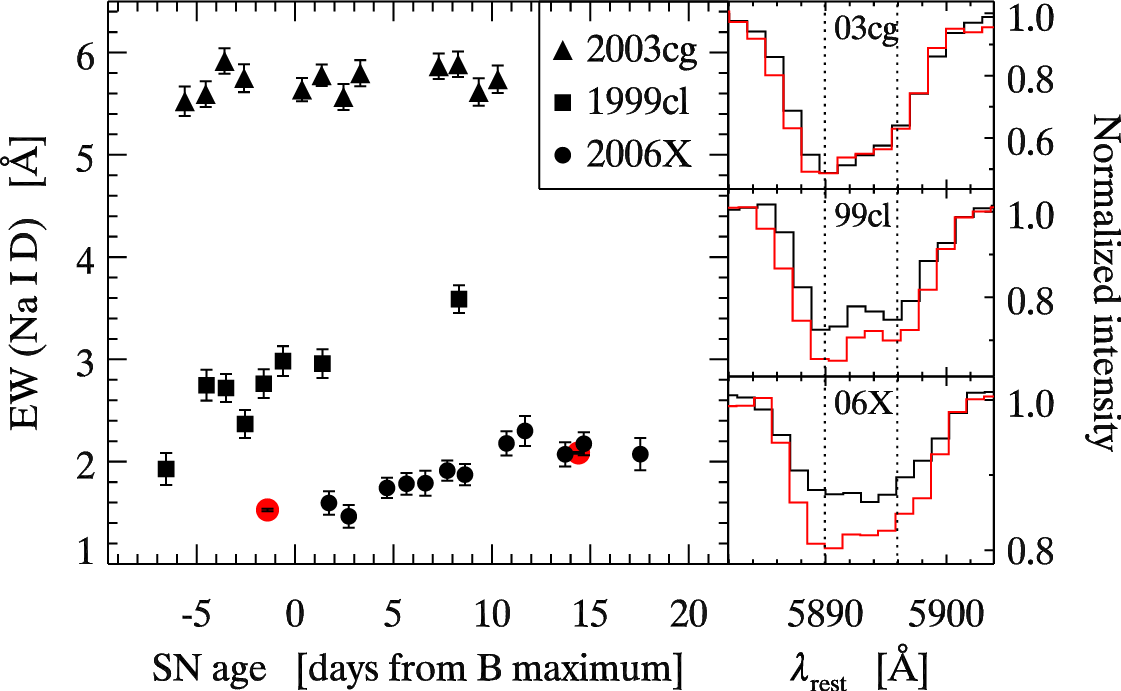}
\caption{\label{fig:na1d}
{\it Left:}
Time-evolution of the equivalent width of the \naoned\ doublet for the
three most highly-reddened \sneia\ in our sample: SNe~2003cg
({\it filled triangles}), 1999cl ({\it filled squares}), and 2006X
({\it filled circles}). Both SNe~1999cl and 2006X exhibit a variable
\naoned\ EW, while that of SN~2003cg remains constant over time. The larger
filled circles at $-2$\,d and +14\,d correspond to EW measurements on
high-resolution (${\rm FWHM}\approx7$\,\kms, or $\sim0.14$\,\AA, at
\naoned) VLT+UVES spectra of SN~2006X published by 
\cite{Patat/etal:2007a}. 
{\it Right:}
Normalized \naoned\ profiles for SN~2003cg ({\it top}), SN~1999cl
({\it middle}) and SN~2006X ({\it bottom}), plotted in rest-frame
wavelength. The black and red lines correspond to the smallest and
largest EW, respectively. Note the difference in ordinate range,
decreasing from top to bottom. The vertical dotted lines indicate the
wavelength positions of the individual D$_1$ and D$_2$ lines.
{\it [See the electronic version of the Journal for a color version of
    this figure.]}
}
\end{figure*}

Using a $\chi^2$ test for a constant EW model, we find that 
two \sneia\ in our sample display variable \naoned\ features:
SNe~1999cl and 2006X. 
While variable sodium features have already
been detected in high-resolution spectra of SN~2006X
\citep{Patat/etal:2007a}, their detection in our data gives us
confidence in our methods and their detection in SN~1999cl constitutes
a new result. 
The $\chi^2$ per degree of freedom (dof) for
the constant EW model is 11.23 for SN~1999cl and 4.47 for SN~2006X
(Table~\ref{tab:ewebv}). We show the EW variation for both objects in
Fig.~\ref{fig:na1d}. All other $\chi^2/{\rm dof}$ are $\lesssim 1.3$.
Interestingly, both SNe~1999cl and 2006X are amongst the three most
highly-reddened objects in our sample (host-galaxy reddening
$E(B-V)_{\rm host}>1$\,mag; see \S~\ref{sect:ewebv}). The other
highly-reddened object, SN~2003cg, 
shows no such variation in the \naoned\ EW ($\chi^2/{\rm dof}=1.07$
for a constant EW model; Table~\ref{tab:ewebv}). The EW variation for
this supernova is also shown in Fig.~\ref{fig:na1d}. The implications
of this result are discussed at length in \S~\ref{sect:disc}.

\begin{figure}
\epsscale{1.1}
\plotone{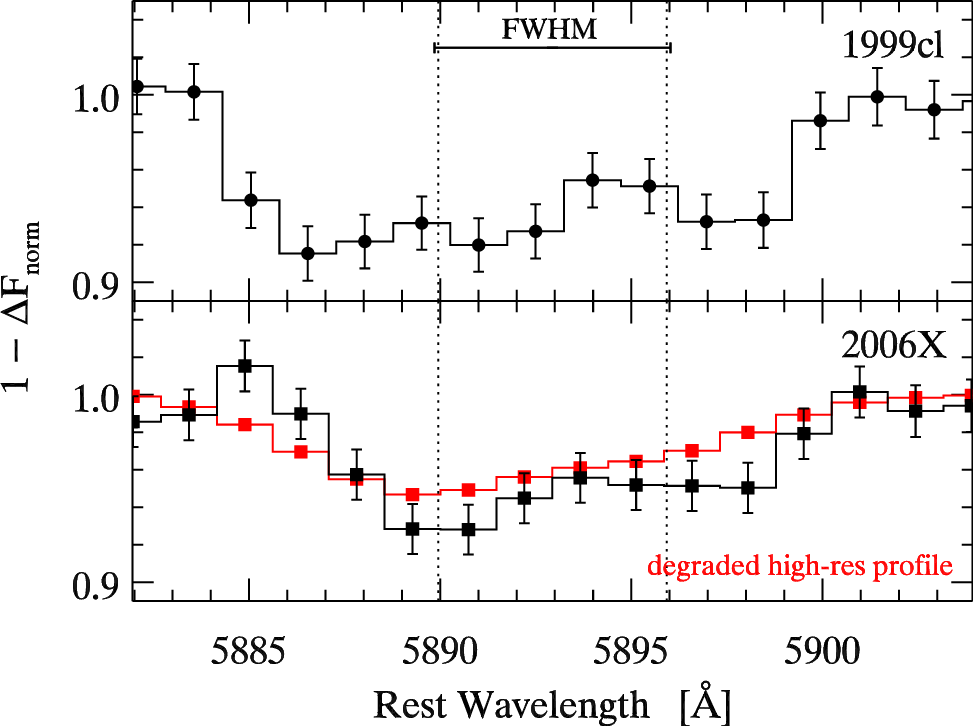}
\caption{\label{fig:nadiff}
One minus the difference between the normalized \naoned\ profiles with
the smallest and largest EW, respectively, for SN~1999cl ({\it top})
and SN~2006X ({\it bottom}). The red line in the lower panel
corresponds to the variation for SN~2006X inferred by degrading the
resolution of the high-resolution spectra published by
\cite{Patat/etal:2007a}. The vertical dotted lines indicate the
wavelength positions of the individual D$_1$ and D$_2$ lines.
The horizontal line in the upper panel shows the size of one
the spectral resolution (${\rm FWHM}\approx6$\,\AA).
{\it [See the electronic version of the Journal for a color version of
    this figure.]}
}
\epsscale{1.0}
\end{figure}

\begin{deluxetable*}{lcccccccc}   
\tabletypesize{\scriptsize}
\tablewidth{0pt}
\tablecaption{Reddening and \naoned\ EW measurements\label{tab:ewebv}}
\tablehead{
\colhead{SN} &
\colhead{$E(B-V)_{\rm host}$} &
\colhead{EW} &
\colhead{$\Delta{\rm EW}$} &
\colhead{$N_\sigma$} &
\colhead{$\chi^2/{\rm dof}$} &
\colhead{dof} &
\colhead{$t_{\rm min}$} &
\colhead{$\Delta t$} \\
 & \colhead{(mag)} & \colhead{(\AA)} & \colhead{(\AA)} & & (const.) & & (d) & (d) \\
\colhead{(1)} & \colhead{(2)} &\colhead{(3)} &\colhead{(4)}
&\colhead{(5)} &\colhead{(6)} &\colhead{(7)} &\colhead{(8)}
&\colhead{(9)} 
}
\startdata
1994D  & 0.04 (0.01) & 0.20$_{-0.06}^{+0.34}$ & 0.40 (0.22) &  1.8 &  1.01 &  4 &  $-$3.6 &  46.8 \\
1994ae & 0.02 (0.01) & 0.42$_{-0.12}^{+0.11}$ & 0.22 (0.11) &  2.0 &  0.87 &  9 &    +1.4 &  28.8 \\
1995al & 0.06 (0.02) & 1.13$_{-0.22}^{+0.10}$ & 0.32 (0.17) &  1.9 &  0.95 &  5 &    +7.0 &  13.0 \\
1997bp & 0.19 (0.03) & 1.85$_{-0.23}^{+0.29}$ & 0.52 (0.29) &  1.8 &  0.96 &  7 &  $-$1.1 &  28.6 \\
1997br & 0.27 (0.04) & 0.54$_{-0.10}^{+0.14}$ & 0.24 (0.16) &  1.5 &  0.79 &  4 &  $-$7.4 &  28.7 \\
1997do & 0.10 (0.03) & 0.32$_{-0.10}^{+0.13}$ & 0.23 (0.20) &  1.1 &  0.56 &  4 &  $-$6.1 &  27.5 \\
1997dt & 0.61 (0.07) & 1.16$_{-0.16}^{+0.27}$ & 0.43 (0.30) &  1.4 &  0.89 &  4 &  $-$9.8 &  10.8 \\
1998aq & 0.01 (0.01) & 0.30$_{-0.14}^{+0.09}$ & 0.23 (0.11) &  2.2 &  0.85 & 13 &  $-$9.0 &  40.8 \\
1998bu & 0.34 (0.04) & 0.27$_{-0.15}^{+0.17}$ & 0.32 (0.16) &  2.0 &  1.07 & 24 &  $-$2.7 &  59.8 \\
1998dh & 0.17 (0.02) & 0.37$_{-0.12}^{+0.16}$ & 0.28 (0.14) &  2.0 &  0.85 &  6 &  $-$8.8 &  54.4 \\
1998dm & 0.34 (0.05) & 0.54$_{-0.26}^{+0.18}$ & 0.44 (0.27) &  1.6 &  0.81 &  5 &    +5.8 &  18.9 \\
1998es & 0.07 (0.02) & 1.58$_{-0.10}^{+0.18}$ & 0.28 (0.14) &  1.9 &  0.85 &  8 &  $-$6.0 &  30.4 \\
1999cl & 1.20 (0.07) & 2.78$_{-0.85}^{+0.81}$ & 1.66 (0.21) &  8.1 & 11.23 &  7 &  $-$6.6 &  14.9 \\
1999dq & 0.12 (0.03) & 1.25$_{-0.19}^{+0.23}$ & 0.42 (0.20) &  2.1 &  0.68 & 12 &  $-$8.6 &  27.6 \\
2001V  & 0.03 (0.02) & 1.34$_{-0.25}^{+0.29}$ & 0.54 (0.28) &  1.9 &  0.83 & 14 &  $-$9.3 &  33.4 \\
2001en & 0.07 (0.04) & 1.10$_{-0.15}^{+0.21}$ & 0.36 (0.18) &  2.0 &  1.24 &  6 &    +2.7 &  12.6 \\
2001ep & 0.14 (0.04) & 0.69$_{-0.31}^{+0.16}$ & 0.47 (0.27) &  1.7 &  0.74 & 10 &  $-$2.7 &  10.8 \\
2002bo & 0.47 (0.04) & 2.64$_{-0.30}^{+0.26}$ & 0.56 (0.21) &  2.6 &  1.07 & 17 &  $-$7.0 &  28.7 \\
2002cr & 0.09 (0.04) & 0.40$_{-0.17}^{+0.16}$ & 0.33 (0.22) &  1.5 &  0.56 &  4 &  $-$7.4 &  12.8 \\
2002fk & 0.03 (0.02) & 0.20$_{-0.09}^{+0.06}$ & 0.15 (0.09) &  1.8 &  0.75 &  5 &    +3.4 &  34.6 \\
2003cg & 1.06 (0.03) & 5.73$_{-0.20}^{+0.19}$ & 0.39 (0.19) &  2.1 &  1.07 & 11 &  $-$5.6 &  15.9 \\
2003du & 0.01 (0.01) & 0.22$_{-0.07}^{+0.10}$ & 0.17 (0.08) &  2.1 &  1.06 &  6 &    +2.2 &  23.8 \\
2003kf & 0.04 (0.03) & 0.51$_{-0.11}^{+0.20}$ & 0.31 (0.16) &  1.9 &  0.92 & 11 &  $-$6.3 &  48.5 \\
2005am & 0.03 (0.02) & 0.23$_{-0.16}^{+0.07}$ & 0.23 (0.11) &  2.1 &  1.13 &  5 &    +1.4 &  11.8 \\
2005cf & 0.10 (0.04) & 0.22$_{-0.08}^{+0.20}$ & 0.28 (0.15) &  1.9 &  0.97 &  9 &  $-$8.8 &  13.9 \\
2006N  & 0.03 (0.02) & 0.67$_{-0.31}^{+0.34}$ & 0.65 (0.28) &  2.3 &  1.35 &  7 &  $-$2.8 &  11.8 \\
2006X  & 1.47 (0.04) & 1.88$_{-0.42}^{+0.42}$ & 0.83 (0.18) &  4.5 &  4.47 & 11 &    +1.7 &  15.8 \\
2007S  & 0.41 (0.03) & 1.81$_{-0.33}^{+0.26}$ & 0.58 (0.28) &  2.1 &  1.32 &  7 &    +0.0 &  38.4 \\
2007af & 0.14 (0.03) & 0.41$_{-0.17}^{+0.24}$ & 0.41 (0.22) &  1.8 &  0.79 & 19 &  $-$4.7 &  68.4 \\
2007bm & 0.52 (0.04) & 2.24$_{-0.31}^{+0.20}$ & 0.51 (0.22) &  2.3 &  1.21 &  6 &    +1.9 &  29.8 \\
2007ca & 0.30 (0.04) & 1.91$_{-0.46}^{+0.29}$ & 0.75 (0.46) &  1.6 &  1.00 &  3 &  $-$0.1 &  10.0
\enddata
\tablecomments{Col.~(1): SN name; col.~(2): host-galaxy color excess determined
from fits to multi-band optical light curves using the MLCS2k2 code
of \cite{MLCS2k2}; col.~(3): weighted mean EW. The upper and lower
limits correspond to the maximum deviations from the weighted mean;
col.~(4): maximum EW difference. The $1\sigma$ error appears in
between parentheses; col.~(5): $\Delta{\rm EW}$ divided by its
$1\sigma$ error; col.~(6): $\chi^2$ per degree of freedom for a
constant EW fit; col.~(7): number of degrees of freedom (simply the
number of data points minus one); col.~(8): age (in days from $B$-band
maximum light) of the earliest spectrum; col.~(9): age range (in days)
of the spectra used in the fit.} 
\end{deluxetable*}   

To our knowledge there are no (multiepoch) high-resolution
spectra of SN~1999cl to confirm our findings, but the agreement
between our measurements on SN~2006X and those of
\cite{Patat/etal:2007a} illustrates the accuracy of our
measurements and Monte Carlo error estimation 
(see Fig.~\ref{fig:na1d}). 
We have further checked that variations in seeing (and thus in
spectral resolution for an object that does not fill up the 3\arcsec\
slit of the FAST spectrograph) do not correlate with the observed
\naoned\ variations.

Note that there is a significant EW variation inferred from
high-resolution spectra at later ages in SN~2006X
($\Delta {\rm EW} \approx 0.36$\,\AA\ between +14 and +61\,d;
\citealt{Patat/etal:2007a}). However, this variation is below our
typical detection threshold of $\sim0.5$\,\AA, and the data from our
observations at these later ages (when the supernova is fainter) do
not generally meet our S/N requirements (see \S~\ref{sect:mc}).

The maximum difference in the total \naoned\ EW is $1.66\pm0.21$\,\AA\
for SN~1999cl (i.e. 8.1$\sigma$ different from zero) and
$0.83\pm0.18$\,\AA\ for SN~2006X (4.5$\sigma$). Without 
high-resolution spectra of SN~1999cl, we cannot place useful
constraints on the change in \naoned\ column density this would
represent. We note that the EW increases with time, and that the
timescale for this variation is $\sim10$~d, for both objects (although
this is a lower limit for SN~1999cl due to lack of suitable
data). The EW variation for SN~2006X between the extrema EW
values (i.e. between +3\,d and +12\,d past maximum) is consistent
with a linear variation, with $\chi^2/{\rm dof}<1$ for a slope of
$\sim0.08$\,\AA\,d$^{-1}$. This is not the case for SN~1999cl, with
$\chi^2/{\rm dof}>2$ for a linear fit. It is difficult to estimate the
significance of the structure seen around $-4$\,d for SN~1999cl with
the available data.

We can examine more closely the \naoned\ profiles in Fig.~\ref{fig:na1d}
to study the nature of the variation seen in SN~1999cl, and in
particular whether it could be different from that seen in
SN~2006X. The high-resolution data for
SN~2006X presented by \cite{Patat/etal:2007a} show that the variation
entirely stems from weaker components blueshifted by $\sim50$\,\kms\
with respect to the main saturated \naoned\ components. Despite the
lower resolution of our data, the \naoned\ profile with the largest EW
(plotted in red in Fig.~\ref{fig:na1d}) does indeed appear skewed
towards bluer wavelengths. This is more clearly visible in the lower
panel of Fig.~\ref{fig:nadiff} where we plot one minus the difference
between the \naoned\ profiles with the smallest and largest EW ({\it
black line}). Degrading the resolution of the high-resolution spectra
of SN~2006X presented by \cite{Patat/etal:2007a} reveals a profile
also skewed towards the blue ({\it red line}).

In contrast, the variation observed in SN~1999cl seems to
affect equally all \naoned\ components between $\sim5885$\,\AA\ and
$\sim5898$\,\AA\ (Fig.~\ref{fig:nadiff}; {\it  top panel}). The
largest variation is observed blueward of the \naonedone\ line (around
$\sim 5886$\,\AA, corresponding to a blueshift of 
$\sim200$\,\kms). However, given the low resolution of our data (${\rm
  FWHM} \approx 6$\,\AA, or $\sim4$ pixels), it is difficult to gauge
the significance of the variation that is not centered at the rest
wavelengths of the \naoned\ lines. Still, if all \naoned\ components
(including blueshifted ones, if present) vary by a similar amount,
this would suggest the variation is associated in part with the same
material where the bulk of the reddening occurs, unlike SN~2006X where
the weaker variable \naoned\ components are not associated with the
material causing the large reddening towards this object (see also
\S~\ref{sect:red}).

Both SNe~1999cl and 2006X are part of the subgroup of \sneia\ with
high velocity gradients in the \sitwo\,\l6355 absorption (HVG; see
\citealt{Benetti/etal:2005}). A larger sample is needed to confirm the
suggestion by \cite{Simon/etal:2007} that the variability is
preferentially associated with the HVG subgroup. We note that our
sample includes several other HVG \sneia\ (e.g., SN~2002bo) that do
not display variable \naoned\ lines.

Our measurements also show that not all blueshifted \naoned\
components in \sneia\ are variable: \cite{Patat/etal:2007a} noted that
a high-resolution spectrum of SN~1998es around maximum light had
similar \naoned\ line profiles as SN~2006X around the same age, but we
see no evidence for variability (${\rm EW} =
1.6^{+0.18}_{-0.10}$\,\AA; see
Table~\ref{tab:ewebv}). \cite{Patat/etal:2007a} also noted similar
profiles in the overluminous SN~1991T, but this object is not part of
the sample presented in this paper.

\begin{figure*}   
\epsscale{0.9}
\plotone{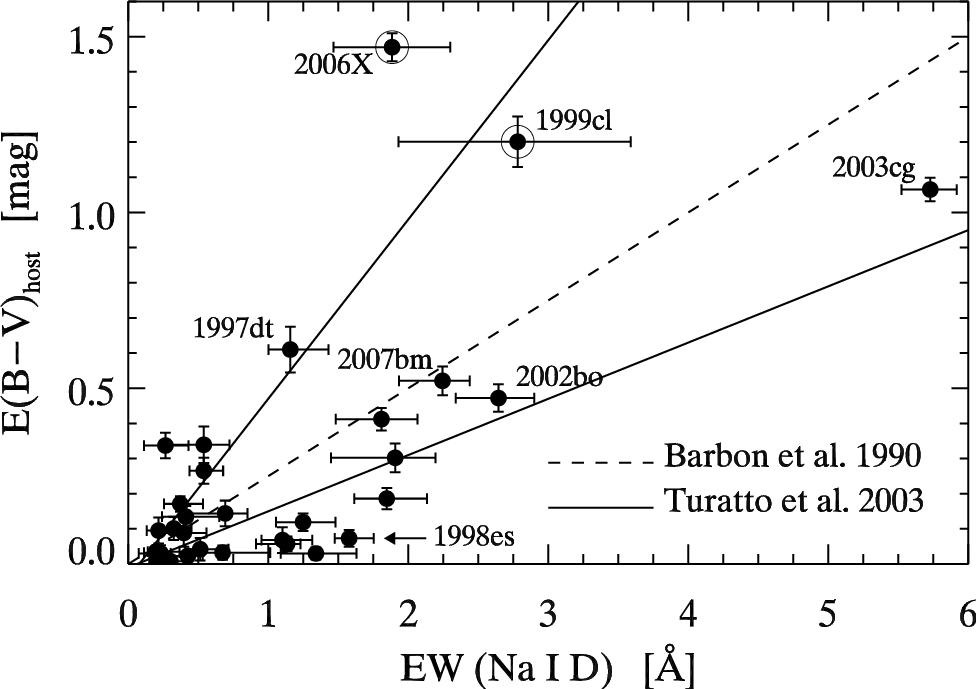}
\caption{\label{fig:ewebv}
Host-galaxy color excess versus equivalent width of the \naoned\
doublet for the 31 \sneia\ in our sample. 
Also shown are the empirical relations of \cite{Barbon/etal:1990} and
\cite{Turatto/Benetti/Cappellaro:2003}, also derived using \snia\ data.
}
\epsscale{1.0}
\end{figure*}   

\subsection{The relation between \naoned\ equivalent width and
  host-galaxy reddening}\label{sect:ewebv}

With \naoned\ equivalent-width measurements for 31 \sneia\ we can
independently investigate its relation with host-galaxy color excess
$E(B-V)_{\rm host}$, as has already been done by several authors using
supernova spectra
\citep{Barbon/etal:1990,Richmond/etal:1994,Turatto/Benetti/Cappellaro:2003}.
In particular, we can test whether the \naoned\ EW can be used to
estimate the host-galaxy reddening for objects with limited photometric data.

The results are displayed in Fig.~\ref{fig:ewebv} and included in
Table~\ref{tab:ewebv}. Here the EW values are reported as the weighted
mean of all measurements for a given supernova, while the error bars
indicate the maximum deviations from the mean (this explains the large
error associated with SN~1999cl, due to intrinsic variations rather
than random error). The color excess in the host galaxy was determined
from fits to multi-band optical light curves using the MLCS2k2 code of 
\cite{MLCS2k2}, corrected for the Galactic reddening
in the line-of-sight using the dust maps of \cite{SFD98}. Over one
half of these light curves are part of a 
new sample that will be published soon by the CfA Supernova Group
\citep{Hicken/etal:2009a}.

While there is a general trend of higher color excess corresponding to
higher \naoned\ EW (such that $E(B-V)_{\rm host} > 0.5$\,mag for ${\rm EW} >
2$\,\AA), the scatter in $E(B-V)_{\rm host}$ at a given EW is considerable. At
${\rm EW}\approx2$\,\AA\ for instance, $E(B-V)_{\rm host}$ ranges from $\sim0.2$
to $\sim1.5$\,mag. Any empirical relation based on \snia\ data can at
best provide an approximate lower limit on $E(B-V)_{\rm host}$. The points do
appear to cluster around the bivariate relation of
\cite{Turatto/Benetti/Cappellaro:2003}, albeit with a larger
spread. The origin of the scatter could be due to a (global) variation
in the dust-to-gas ratio in the host-galaxy interstellar medium,
associated with variations in gas metallicity and dust composition
\citep[e.g.,][]{Draine/etal:2007}. Effects local to the SN circumstellar
environment in the form of light echoes have also been found to impact
the reddening derived assuming a standard color evolution for \sneia\
\citep{WangL:2005,Patat/etal:2006}.

We note that the relation of color excess with EW is expected to be
non-linear for a fixed dust-to-gas ratio because of line saturation
effects at high EW, as observed by
\cite{Munari/Zwitter:1997} in Galactic O and B stars. Linearity at
high EW could hold if the extinction takes place in a
number of clouds at different velocities, each producing unsaturated
features.


\section{Circumstellar or interstellar origin for the variable
  \naoned\ lines?}\label{sect:disc}

\subsection{Conflicting interpretations of the \naoned\ variability}

In SN~2006X, the variability of blueshifted \naoned\ lines was
interpreted by \cite{Patat/etal:2007a} as the result of changing
ionization conditions in circumstellar 
material within the progenitor system due to UV radiation from the 
supernova. The various sub-components in \naoned\ then originate in
the pre-SN variable wind of an early red giant companion, or from
remnant shells ejected in successive nova explosions
\citep{Patat/etal:2007a}. In the latter case, an asymmetric wind from
the companion is needed to decelerate the material ejected during nova
outbursts to match the observed $\lesssim100$\,\kms\ blueshifts of the
\naoned\ components. In principle, the low fraction of objects for
which we detect variable \naoned\ lines (2/31, i.e. $\sim6$\%) could
be due to viewing angle effects (see \citealt{Patat/etal:2007b}).

In both SNe~1999cl and 2006X, the \naoned\ EW increases with age, as
expected from Na II$\rightarrow$I recombination. The variation occurs on
a similar $\sim10$\,d timescale, although this is a lower limit for
SN~1999cl due to lack of data (see Fig.~\ref{fig:na1d}). At
first sight, the \naoned\ variability seems consistent with the
interpretation of \cite{Patat/etal:2007a}, that it is
associated with CSM in the \snia\ progenitor system.

In a recent paper, \cite{Chugai:2008a} argues that 
the blueshifted \naoned\ components in SN~2006X cannot originate
within CSM material from a red giant wind.
The optical depth in \naoned\ would render it undetectable (even
in high-resolution data), while that in CSM lines of \catwo\,H\&K
would be sufficient to detect them. The observed lack of variability
in these calcium lines is central to the argument of
\cite{Patat/etal:2007a} that the features observed in \naoned\ cannot
be due to intervening clouds of interstellar material that are not
associated with the supernova progenitor system. In contrast,
\cite{Chugai:2008a} use the same observational fact about \catwo\ to
set an upper limit on the (spherically symmetric) wind from the RG
companion. Unfortunately, we cannot study the variability in
\catwo\,H\&K lines with our data due to lower S/N in the blue portion
of our spectra. 

The variable \naoned\ lines observed in SN~2006X are
interpreted by \cite{Chugai:2008a} as being due to clusters of
interstellar clouds with different calcium abundances. Whether or not
this is a plausible explanation is beyond the scope of this paper, but
a direct consequence of this (noted by \citealt{Chugai:2008a}) is that
the variability in \naoned\ should be preferentially associated with 
highly-reddened \sneia. This is certainly the case in our sample:
two of the three \sneia\ with $E(B-V)_{\rm host} > 1$\,mag exhibit variable
\naoned. If these time-variant features were randomly associated to
all \sneia, independent of their reddening, the probability of
finding such features only in two highly reddened Ia is very low
($\sim0.6$\%). This fact alone is not proof that the variability
is related to interstellar absorption, but is suggestive enough to
warrant further scrutiny once multiepoch high-resolution spectroscopy
for a larger sample is available.

\subsection{The Reddening Towards SNe~1999cl and 2006X}\label{sect:red}

Both SNe~1999cl and 2006X were found to be reddened by dust
with unusual values of $R_V$, the ratio of total to selective
extinction ($R_V\approx1.6$ for SN~1999cl,
\citealt{Krisciunas/etal:2006}; $R_V\approx1.5$ for SN~2006X,
\citealt{WangX/etal:2008a}), to be compared with $R_V\approx3.1$ for
the average value in the Milky-Way
\citep[e.g.,][]{CCM89}. Spectropolarimetry of SN~2006X has also
shown that the dust producing the polarization is different from the
typical Milky Way mixture \citep{Patat/etal:2008}.
Studies of larger samples also conclude that dust in
the host galaxies of \sneia\ is characterized by low values of $R_V$
(e.g., \citealt{Conley/etal:2007,Hicken/etal:2009a}).

However, not all \sneia\ reddened by non-standard dust display
variable \naoned\ lines: SN~2003cg has a color excess $E(B-V)_{\rm
  host}\approx1.1$\,mag and $R_V\approx1.8$
\citep{Elias-Rosa/etal:2006}, yet its \naoned\ equivalent width
remains constant from one week before to two weeks past maximum light
(mean ${\rm EW} \approx 5.7\pm0.2$\,\AA; see Table~\ref{tab:ewebv}).
This suggests the variability is not systematically related to
non-standard dust. If low values of $R_V$ arise in circumstellar
material around \sneia\
\citep[see][]{WangL:2005,Patat/etal:2006,Goobar:2008}, this in turn
implies that the \naoned\ variability does not necessarily result from
effects local to the SN environment.

\cite{Patat/etal:2007a} attribute the large reddening towards SN~2006X
to a dense molecular cloud, causing the strongly saturated \naoned\ lines
and a number of weaker features (CN, CH, \caone, diffuse interstellar
bands, or DIB) at the same heliocentric velocity \citep{Cox/Patat:2008}.
The column density
inferred from the strongly saturated \naoned\ lines is $\log{N({\textrm
\naone})} \approx 14.3$, which is two orders of magnitude larger than
that of the most intense time-variant features ($\log{N}\approx 12$;
\citealt{Patat/etal:2007a}). In this picture, the reddening produced by the
material where these variable features arise, if any, is
negligible. Additional evidence comes from the detection of a light
echo associated with SN~2006X
\citep{WangX/etal:2008b,Crotts/Yourdon:2008}.
\cite{Crotts/Yourdon:2008} present compelling evidence that this echo
is generated in a dust sheet $\sim26$\,pc in front of the supernova,
presumably where the bulk of the reddening occurs, and hence is not
associated with circumstellar material. 

In SN~1999cl, the variation appears to affect all \naoned\
components, including some with blueshifts of up to $\sim200$\,\kms
(see Fig.~\ref{fig:nadiff}), although the presence of such blueshifted
components is difficult to confirm given the low resolution of our
spectra (1 pixel corresponds to $\sim75$\,\kms\ at \naoned). 
We also note that we detect the DIB at $\sim6284$\,\AA\ in our
low-resolution spectra of SN~1999cl at the same velocity as the
\naoned\ absorption (the spectra of SN~1999cl have been published by
\citealt{Matheson/etal:2008} and are publicly available at the
CfA Supernova Archive). However, this feature is too weak (${\rm
  EW}\lesssim 0.5$\,\AA) for us to confidently detect any variability.

The fact that we detect variable \naoned\ features
only in the two most highly-reddened \sneia\ in our sample does
suggest that the variability is somehow connected to dusty
environments, and is not necessarily due to circumstellar absorption.

\subsection{Variable Interstellar Lines in Galactic Stars, GRBs, and
  the Type IIn SN~1998S}

Time-variable interstellar absorption lines have been detected in
high-resolution spectra of Galactic stars (see \citealt{Crawford:2003}
and references therein). However, the EW variation is typically
$\lesssim 0.01$\,\AA\ and occurs on the timescale of years to
decades (cf. $\sim1$\,\AA\ over $\sim10$\,d for SNe~1999cl and
2006X). Another important difference is that the variation occurs in
several interstellar lines (e.g., both \naone\ and \catwo;
\citealt{Crawford/etal:2002}), as opposed to \naoned\ only (as
observed by \citealt{Patat/etal:2007a} in SN~2006X).
In most cases, the variation is associated with interstellar
shells and bubbles (e.g., 7 of the 13 variable interstellar sight lines
compiled by \citealt{Crawford:2003} are associated with the Vela
SNR). The timescale of the variation for these stars implies the
existence of structure in the ISM on scales of 10--100\,AU (e.g.,
\citealt{Lauroesch/Meyer/Blades:2000}).

Time-variable interstellar lines have also been observed in other
extragalactic objects. \cite{Hao/etal:2007} report on the detection of
variable \fetwo\ and \mgtwo\ lines in low-resolution spectra of GRB
060206 (although see \citealt{Thoene/etal:2008}), which they interpret
as stemming from insterstellar patches 
of \mgtwo\ clouds of size comparable to the GRB beam size
($\sim10^{16}$\,cm). \cite{D'Elia/etal:2008} have revealed variable
lines of \fetwo\ in high-resolution spectra of GRB 080319B, whose fine
structure was excited through pumping by the GRB UV photons. The
distance to the absorbing material is constrained to be
$\sim$18--34\,kpc away. (Note that the UV flux of a GRB is orders of
magnitude larger than that of a \snia, whose ionizing/exciting
capabilities are rather limited in space). In both cases the variation
occurs over very short timescales (a few hours).

As noted by \cite{Patat/etal:2007a}, the only other supernova with
variable (and blueshifted) \naoned\ lines is the (core-collapse) Type
IIn SN~1998S \citep{Bowen/etal:2000}. In this object, however, the
presence of narrow P-Cygni profiles of H and He at the same velocity
as the variable \naoned\ lines is interpreted as the signature of
outflows from the supergiant progenitor, hence associating the
variability with changes in the CSM
(\citealt{Bowen/etal:2000,Fassia/etal:2001}; see also
\citealt{Fransson/etal:2005}).


\section{Conclusion}\label{sect:ccl}

We detect variable \naoned\ lines in low-resolution spectra
of 2/31 ($\sim6$\,\%) \sneia: SNe~1999cl and 2006X. 
The S/N of our spectroscopic data set is sufficiently high
  (S/N$>50$ per pixel) that we can reliably detect changes of
$\gtrsim0.5$\,\AA\ in the \naoned\ equivalent width.
This is comparable to the total EW variation seen
in SN~2006X between $-2$ and +14\,d past maximum light 
\citep{Patat/etal:2007a}. We conclude that either variable \naoned\
features are not a common property of \sneia, or that the level of the
variation is less on average than observed in SN~2006X.

The limitations of this study due to the low resolution of our spectra
are overcome in part by the larger statistical power of our sample. 

Only a small fraction ($\sim6$\%\ in our sample) of \sneia\ display
variable \naoned\ lines. The two \sneia\ for which we detect variable
\naoned\ (SNe~1999cl and 2006X) are the two most highly reddened
  objects in our sample, suggesting that the variability could be
associated with interstellar absorption.
However, we detect no such variability in the other highly-reddened
SN~2003cg. We cannot exclude with our data the possibility that a
larger fraction of objects display EW variations below our detection
threshold of $\sim0.5$\,\AA.

Whether the material causing the large \naoned\ EW variations is
connected to the \snia\ progenitor system ought to become clear from
multiepoch high-resolution spectroscopy for a larger sample.
The detection of such variable \naoned\ features in an unreddened
\snia\ would reduce the probability that the variability results from
a purely geometrical effect in clouds of interstellar material.
If variable \naoned\ lines remain a rare property amongst
\sneia, this would either indicate that the variability is not
associated with circumstellar material (and hence to the \snia\
progenitor system), or that there is a preferred CSM geometry for
detecting the variability. It is premature with the current
available data to place firm constraints on the nature of Type Ia
supernova progenitor systems based on the presence or absence of
variable \naoned\ lines alone.


\acknowledgments
It is a pleasure to thank all observers on the FLWO 1.5\,m telescope,
and in particular Perry Berlind and Mike Calkins, for their continued
support and dedication in obtaining data for the CfA Supernova Program.
We further thank Gast\'on Folatelli and Mario Hamuy of the Carnegie Supernova
Project for communicating their $\Delta m_{15}(B)$ value for SN~2005am
ahead of publication. Similar thanks go to Weidong Li for sharing the
results of his MLCS2k2 fit to SN~2005am. We also acknowledge useful
discussions with Joe Liske, Josh Simon, Xiaofeng Wang, and David Weinberg. 
MM acknowledges support from a Miller Institute Research Fellowship 
during the time in which part of this work was completed.
We are grateful to the anonymous referee for prompting us to
investigate possible differences in the nature of the \naoned\
variability seen in SN~1999cl and SN~2006X.
Support for supernova research at Harvard University, including the
CfA Supernova Archive, is provided in part by NSF grant AST
06-06772.

{\it Facilities:} 
\facility{FLWO:1.5m (FAST)}


\bibliographystyle{apj}
\bibliography{ms_orig}

\end{document}


\title{Erratum: ``A Second Case of Variable N\lowercase{a}~I~D Lines in a
  Highly-Reddened Type I\lowercase{a} Supernova'' (2009, \apj, 693, 207)} 

\author{St\'ephane Blondin,\altaffilmark{1,2,*}
  Jos\'e L. Prieto,\altaffilmark{3,$\dag$}
  Ferdinando Patat,\altaffilmark{2}
  Peter Challis,\altaffilmark{1}
  Malcolm Hicken,\altaffilmark{1}
  Robert P. Kirshner,\altaffilmark{1,$\ddag$}
  Thomas Matheson,\altaffilmark{4}
  Maryam Modjaz\altaffilmark{5,\S}
}

\altaffiltext{1}{Harvard-Smithsonian Center for Astrophysics, 60 Garden
  Street, Cambridge, MA 02138}
\altaffiltext{2}{European Southern Observatory, Karl-Schwarzschild-Strasse
  2, D-85748 Garching, Germany}
\altaffiltext{3}{Dept. of  Astronomy, The Ohio State
  University, 140 W. 18th Ave., Columbus, OH 43210}
\altaffiltext{4}{National Optical Astronomy
  Observatory, 950 North Cherry Avenue, Tucson, AZ 85719-4933}
\altaffiltext{5}{Department of Astronomy, University of
  California, Berkeley, CA 94720-3411}

\altaffiltext{*}{Now at: Aix Marseille Univ, CNRS, LAM, Laboratoire
  d'Astrophysique de Marseille, Marseille, France}
\altaffiltext{$\dag$}{Now at: N\'ucleo  de  Astronom\'ia  de  la  Facultad  de
  Ingenier\'ia, Universidad Diego Portales,  Av. Ej\'ercito  441,
  Santiago, Chile; Millennium Institute of Astrophysics, Santiago,
  Chile}
\altaffiltext{$\ddag$}{Now also at: Gordon and Betty Moore Foundation, 1661 Page
  Mill Road, Palo Alto, CA 94304, USA}
\altaffiltext{\S}{Now at: Center for Cosmology and Particle Physics, New
   York University, 4 Washington Place, New York, NY 10003, USA}

\section{No variable \naoned\ lines in SN~1999\lowercase{cl}}

\begin{figure*}
\plottwo{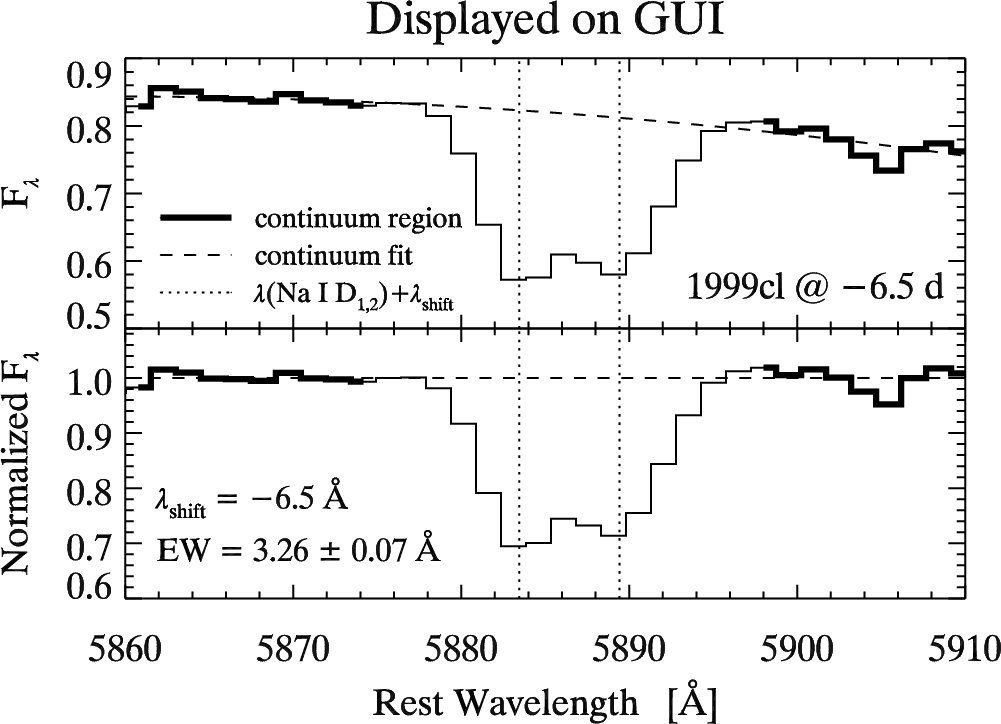}{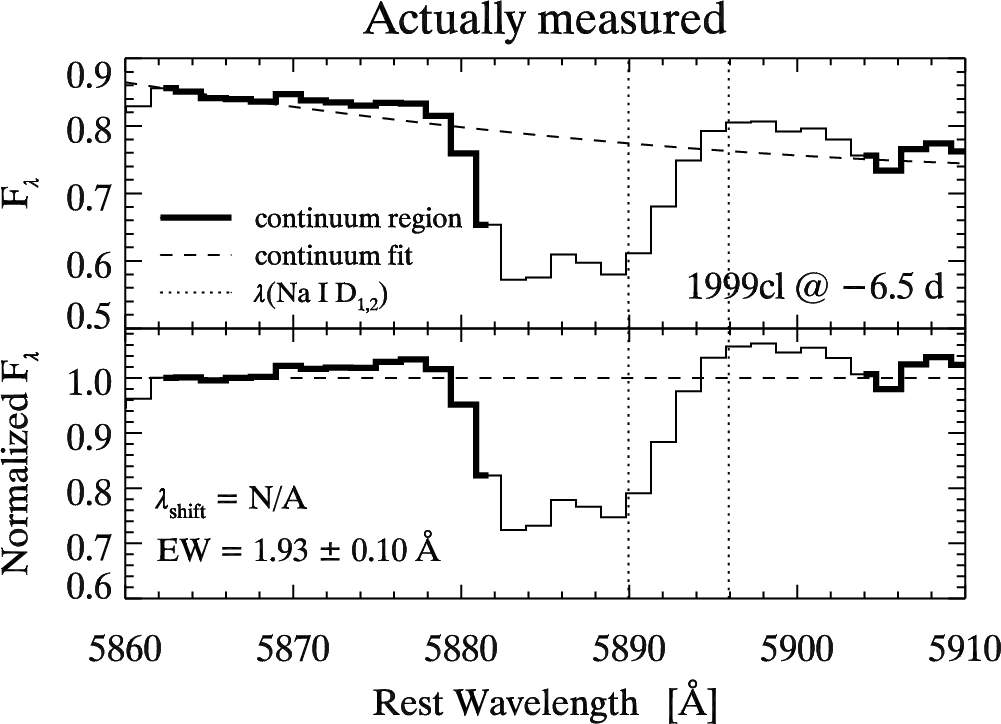}
\caption{\label{fig:wshift}
Impact of the observed wavelength shift of the \naoned\ profile on the
EW measurement, illustrated using the SN~1999cl spectrum at $-6.5$\,d from maximum light. The
left panel reproduces what was displayed on the graphical interface
used to determine the continuum-fit regions on either side of the
\naoned\ profile (thick black lines), with the rest-wavelength
locations of the D$_2$ and D$_1$ lines shifted by $\lambda_{\rm
  shift}=-6.5$\,\AA\ (dotted lines) with respect to their true
locations (5889.95 and 5895.92\,\AA, respectively). Division by the
continuum fit (dashed line) yields a normalized profile (lower panel)
appropriate for computing the profile EW, here determined to be
$3.26\pm0.07$\,\AA.  However, in our revised implementation of the EW
computation the wavelength ranges of the continuum-fit regions were
passed to a separate function where the wavelength shift was
accidentally ignored (i.e., $\lambda_{\rm shift}=0$). Since the
continuum wavelength ranges were determined with respect to the
rest-wavelength locations of the D$_2$ and D$_1$ lines, this resulted
in an erroneous continuum fit and subsequent underestimate of the true
EW ($1.93\pm0.10$\,\AA\ instead of $3.26\pm0.07$\,\AA). SN~1999cl is
the only object affected by this analysis error, which thus went
unnoticed in our various validation tests based on SN~2006X.
}
\end{figure*}

\begin{figure*}
\plottwo{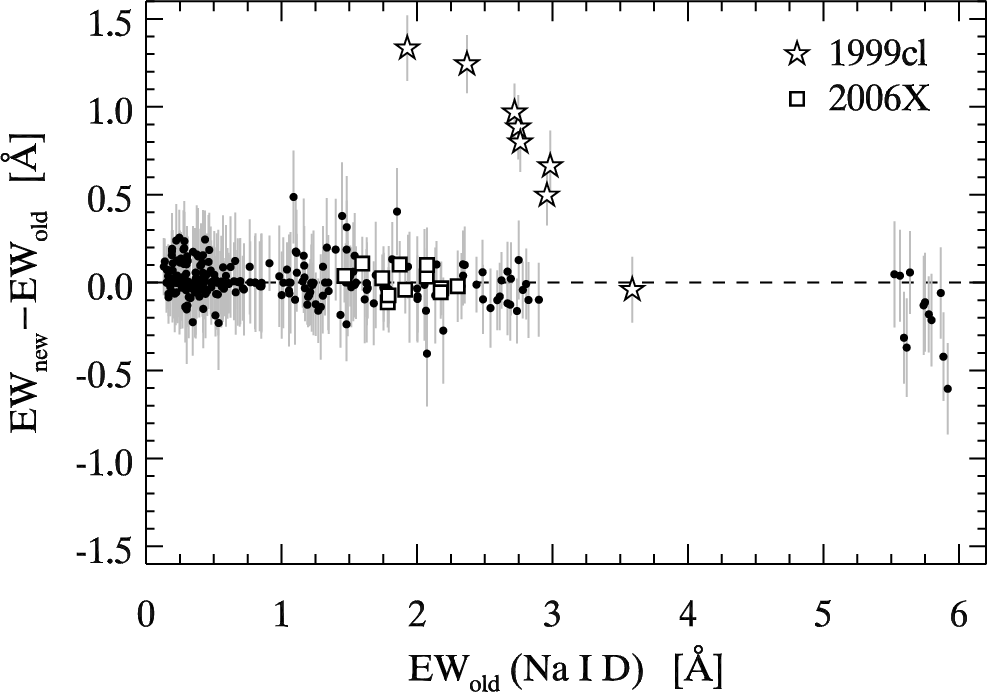}{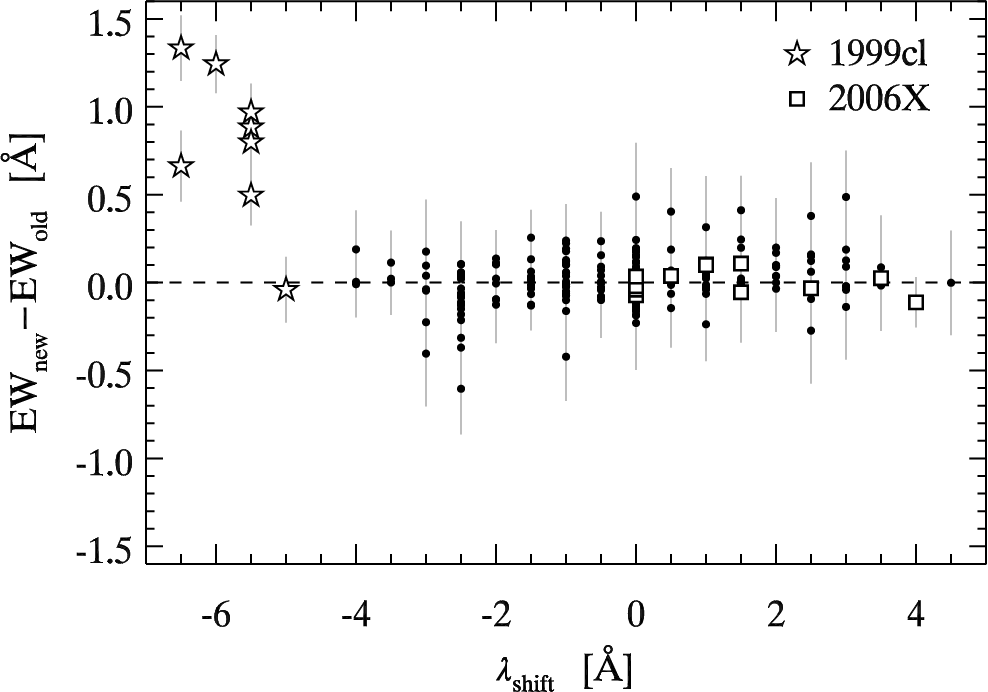}
\caption{\label{fig:comp_meas}
{\it Left:} Difference between our revised (EW$_{\rm
    new}$) and original (EW$_{\rm old}$) \naoned\ EW
measurements {\it vs.}  our original measurements. Only SN~1999cl is
systematically and largely affected (open stars), with typical
differences in excess of 0.5\,\AA\ at the 3$\sigma$ level. Excluding
SN~1999cl, the RMS of all other residuals is only $\sim0.13$\,\AA. The
RMS residual for SN~2006X is even lower, at $\sim 0.07$\,\AA\ (open squares).  {\it
  Right:} Difference between our revised and
original \naoned\ EW measurements 
{\it vs.}  the measured wavelength shift of the \naoned\ profile
($\lambda_{\rm shift}$), manually adjusted in 0.5\,\AA\ steps. Only
SN~1999cl (open stars) displays absolute shifts at the $\gtrsim 5$\,\AA\ level,
while measurements corresponding to $|\lambda_{\rm shift}| \lesssim
4$\,\AA\ are not significantly affected (including SN~2006X, open
squares).
}
\end{figure*}

The large variation in the \naoned\ equivalent width (EW) observed in
the Type Ia SN~1999cl \citep{Blondin/etal:2009}, $\Delta\mathrm{EW} =
1.66\pm0.21$\,\AA, results in fact from a 
measurement error. The origin of this error was traced back to
observed wavelength shifts of the \naoned\ profile with respect to its
expected restframe location (5889.95 and 5895.92\,\AA\ for
  the D$_2$ and D$_1$ lines, respectively), which were not properly
taken into
account in a revised implementation of our EW computation (albeit
correctly displayed on the graphical interface developed for these
measurements; see Fig.~\ref{fig:wshift}). We revised {\it
    all} the measurements in the original paper, and can confirm that
SN~1999cl is the only object 
of our sample affected by this error (see
Fig.~\ref{fig:comp_meas}). In particular,
measurements for SN~2006X are not affected when properly
  taking into account these wavelength shifts (see open squares in
  Fig.~\ref{fig:comp_meas}). Since 
our measurement routines
were validated based on SN~2006X, for which high-resolution VLT+UVES data were
available \citep{Patat/etal:2007a}, and for which our own EW
determinations on lower-resolution FLWO 1.5\,m+FAST data yielded
consistent results,
this error went unnoticed.

SN~1999cl displays the largest wavelength shifts of the
\naoned\ profile of all the \sneia\ in our sample, with typical shifts
of approximately $-6$\,\AA, corresponding to a $\sim
300$\,\kms\ velocity blueshift with respect to its expected
restframe location. This velocity shift corresponds to the
difference between the recession velocity of the host-galaxy nucleus
(2338\,\kms\ for NGC 4501; \citealt{UZC}), which we used to
  de-redshift the SN~1999cl spectra, and that derived at the 
location of SN~1999cl from emission lines in the SN spectrum used for
classification purposes by \cite{Garnavich_iauc7198}, namely $cz =
2030$\,\kms. It results from the location of SN~1999cl in a spiral arm
with a blueshifted velocity along the line of sight, as derived from
kinematic maps of NGC 4501 based on H\,{\sc I} emission by
\cite{Chemin/etal:2006}. We note that \cite{Garnavich_iauc7198} had in
fact correctly reported a \naoned\ EW of 0.33\,nm for their first
spectrum of SN~1999cl, consistent with our revised measurement on the
same spectrum.

All the other objects in our sample display significantly
  lower wavelength shifts of the \naoned\ profile ($\lesssim
  4$\,\AA\ in absolute value; see Fig.~\ref{fig:comp_meas}). We have
  not investigated the exact nature of the observed wavelength shifts
  for all the objects in our sample, but simply note that absolute
  shifts at the $\lesssim 4$\,\AA\ level do not appear to affect our
  revised EW measurements, with an RMS difference compared to our
  original measurements of only $\sim 0.13$\,\AA.

Our new measurements show that the EW variation for the
\naoned\ profile in SN~1999cl is significantly lower than reported in
our original paper ($0.43\pm0.14$\,\AA\ cf. $1.66\pm0.21$\,\AA). While
the EW variation remains statistically significant (3.1$\sigma$
different from zero), it is now below the detection threshold of
0.5\,\AA\ derived from the Monte Carlo simulations published in the
original paper (these simulations are unaffected by the aforementioned measurement
error). We also note that the S/N cut based on these same MC
simulations ($\mathrm{S/N} \ge 50$ per pixel) was erroneously applied based on the S/N per \AA\ instead
of per dispersion element (typically $\sim 1.5$\,\AA\ per pixel),
resulting in the exclusion of spectra with $50 \le \mathrm{S/N\ per\
  pixel} \le 50 \times 1.5 = 75$. Our revised measurements thus
typically include more spectra for a given SN than in the original
paper (see Table~\ref{tab:ewebv}).

As a result, SN~1999cl should no longer be considered as an object
displaying variable \naoned\ lines in our study. The fraction of
\sneia\ in our sample displaying \naoned\ lines thus goes from $\sim 6$\%
(2/31) in the original study to $\sim 3$\% (1/31) in the revised
analysis, SN~2006X being the only \snia\ in our sample with
variable \naoned\ lines.

The main conclusion of our paper remains unchanged. Quoting from the
conclusions section in the original paper: ``We conclude that either
variable \naoned\ features are not
a common property of \sneia, or that the level of the variation
is less on average than observed in SN 2006X.'' However, the title
of our paper highlighted the new detection of ``variable \naoned\ lines
in a highly-reddened Type Ia supernova'' (i.e., SN~1999cl), which no
longer holds. 

Our revised measurements affect Figs.~3, 4, and 5 of the original
paper, as well as Table~2. We present revised versions of these in
what follows (Figs.~\ref{fig:na1d}, \ref{fig:nadiff}, \ref{fig:ewebv},
and Table~\ref{tab:ewebv}). In addition, we noticed a small error in
the legend labels of Fig.~2 in the original paper, which is corrected
in the new Fig.~\ref{fig:ewerr} below.

We are indebted to Lingzhi Wang, currently a research scholar
at the Chinese Academy of Sciences South America Center for Astronomy
in Santiago, Chile, for having brought this measurement error to our
attention during the conference ``Supernovae through the ages'' held
on Easter Island, Chile, during August 2016.

\epsscale{1.1}
\epsscale{1.0}

\epsscale{1.1}
\epsscale{1.0}

\epsscale{1.1}
\begin{figure}
\plotone{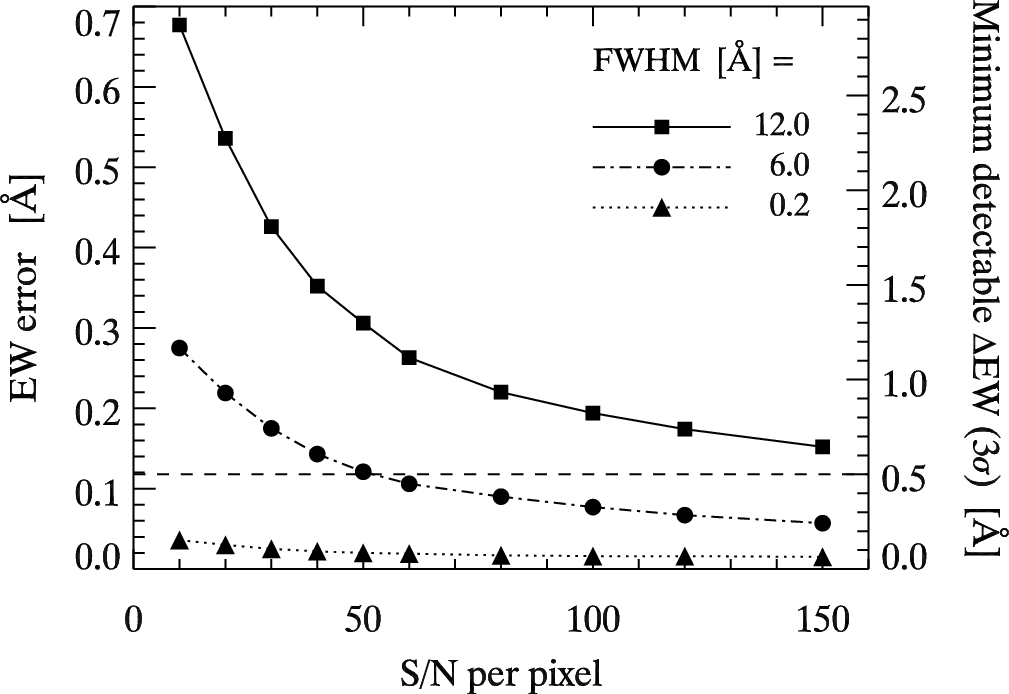}
\caption{\label{fig:ewerr}
{\it [Figure~2 in \cite{Blondin/etal:2009}].}
EW measurement error as a function of S/N for three FWHM
resolutions. The dash-dotted curve corresponds to the 
resolution of our data set (${\rm FWHM}\approx 6$\,\AA). The ordinate
axis on the right indicates the corresponding 3$\sigma$ error on the
difference between EW measurements. The horizontal dashed line
corresponds to a minimum detectable EW difference of 0.5\,\AA.
}
\end{figure}
\epsscale{1.0}

\begin{figure*}
\plotone{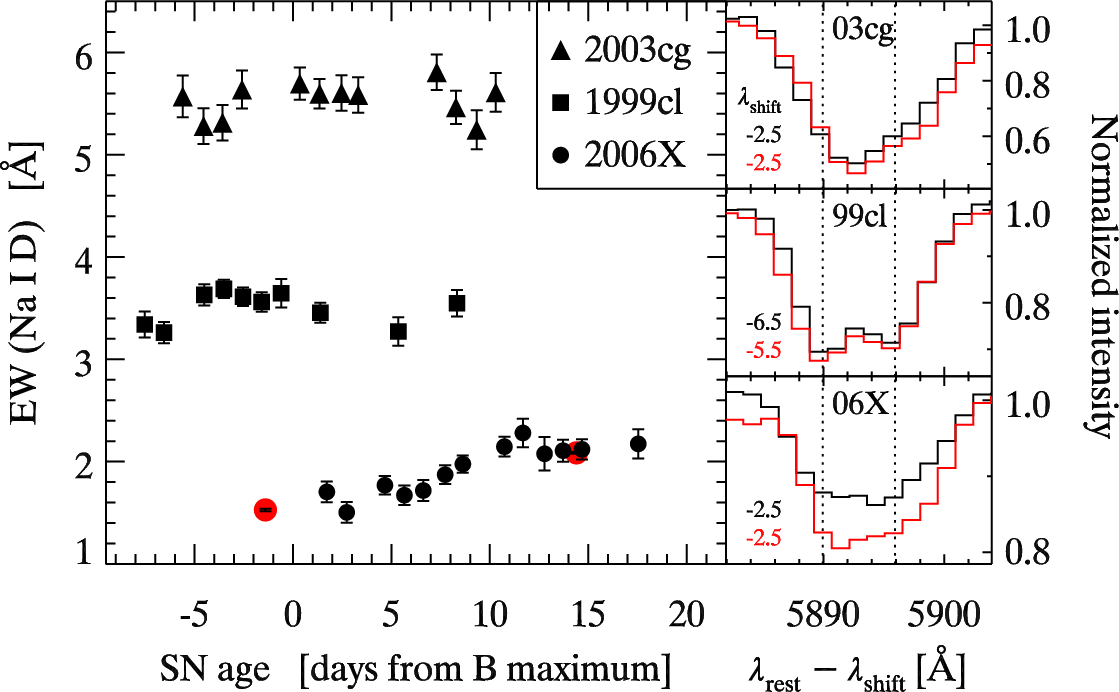}
\caption{\label{fig:na1d}
{\it [Figure~3 in \cite{Blondin/etal:2009}].}
{\it Left:}
Time-evolution of the equivalent width of the \naoned\ doublet for the
three most highly-reddened \sneia\ in our sample: SNe~2003cg
({\it filled triangles}), 1999cl ({\it filled squares}), and 2006X
({\it filled circles}). Only SN~2006X exhibits clear variable
\naoned\ EWs, while those of SN~1999cl and SN~2003cg remain largely
constant over time. The larger 
filled circles at $-2$\,d and +14\,d correspond to EW measurements on
high-resolution (${\rm FWHM}\approx7$\,\kms, or $\sim0.14$\,\AA, at
\naoned) VLT+UVES spectra of SN~2006X published by 
\cite{Patat/etal:2007a}. 
{\it Right:}
Normalized \naoned\ profiles for SN~2003cg ({\it top}), SN~1999cl
({\it middle}) and SN~2006X ({\it bottom}), plotted in rest-frame
wavelength corrected for the shift in wavelength of the
  \naoned\ doublet, $\lambda_{\rm shift}$, whose value is indicated
  for each profile. The black and red lines correspond to the smallest and
largest EW, respectively. Note the difference in ordinate range,
decreasing from top to bottom. The vertical dotted lines indicate the
wavelength positions of the individual D$_2$ and D$_1$ lines.
{\it [See the electronic version of the Journal for a color version of
    this figure.]}
}
\end{figure*}

\begin{figure}
\epsscale{1.1}
\plotone{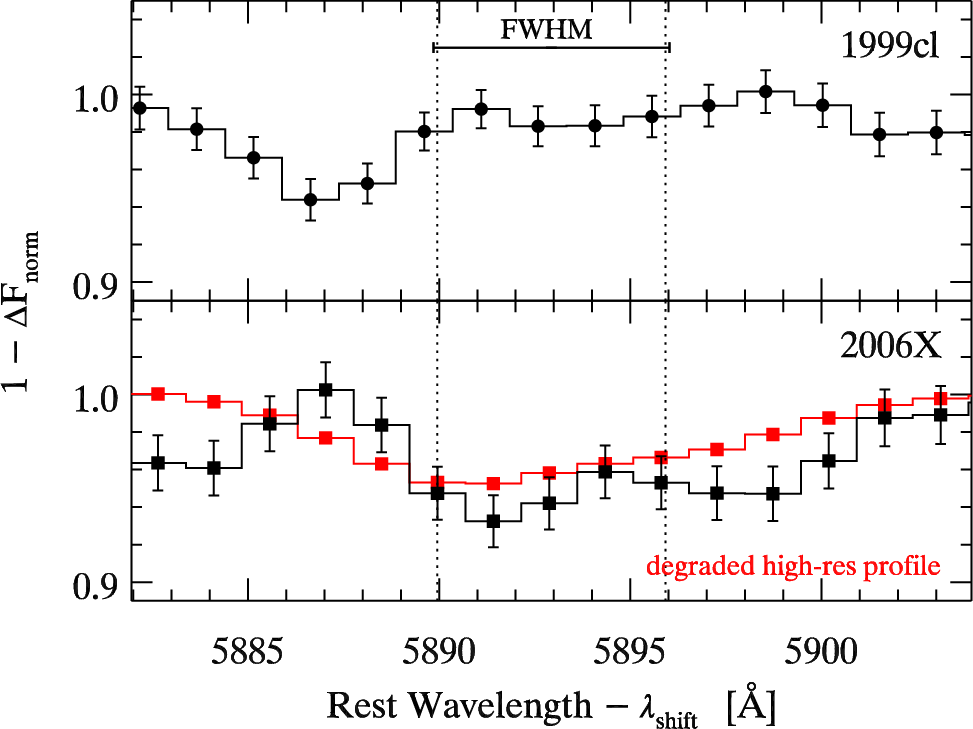}
\caption{\label{fig:nadiff}
{\it [Figure~4 in \cite{Blondin/etal:2009}].}
One minus the difference between the normalized \naoned\ profiles with
the smallest and largest EW, respectively, for SN~1999cl ({\it top})
and SN~2006X ({\it bottom}). The wavelengths have been corrected
  for the shift in wavelength of the 
  \naoned\ doublet, $\lambda_{\rm shift}$ (see Fig.~\ref{fig:na1d}). The red line in the lower panel
corresponds to the variation for SN~2006X inferred by degrading the
resolution of the high-resolution spectra published by
\cite{Patat/etal:2007a}. The vertical dotted lines indicate the
rest wavelength positions of the individual D$_2$ and D$_1$ lines.
The horizontal line in the upper panel shows the size of one spectral
resolution element (${\rm FWHM}\approx6$\,\AA). 
{\it [See the electronic version of the Journal for a color version of
    this figure.]}
}
\epsscale{1.0}
\end{figure}

\begin{figure*}
\plotone{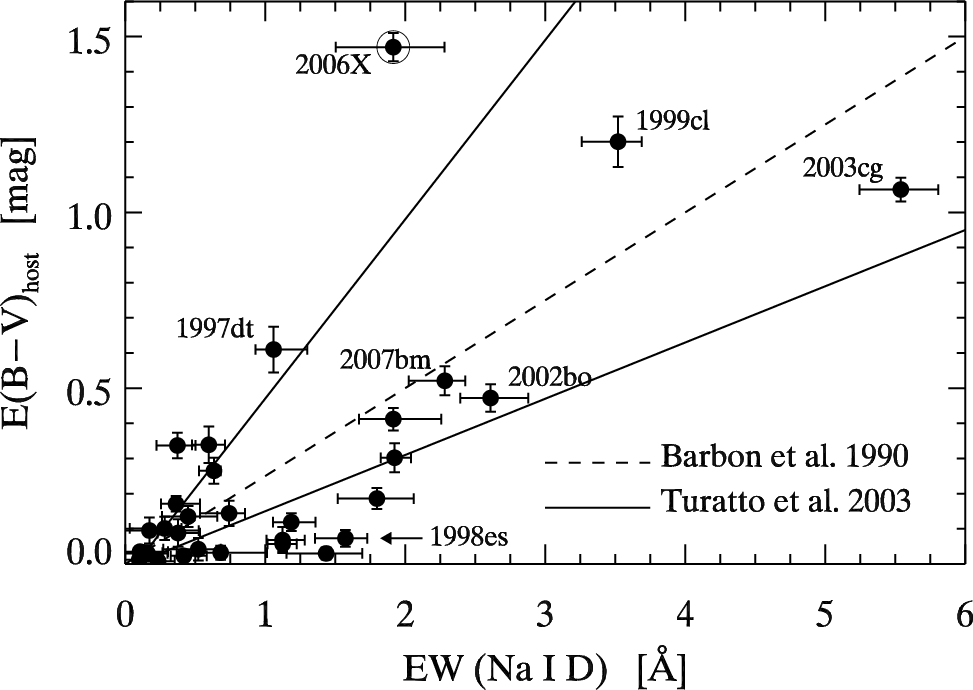}
\caption{\label{fig:ewebv}
{\it [Figure~5 in \cite{Blondin/etal:2009}].}
Host-galaxy color excess versus equivalent width of the \naoned\
doublet for the 31 \sneia\ in our sample.
The error bar for EW(\naoned) corresponds to the full range of
the observed variation.
Also shown are the empirical relations of \cite{Barbon/etal:1990} and
\cite{Turatto/Benetti/Cappellaro:2003}, also derived using \snia\ data.
}
\end{figure*}

\begin{deluxetable}{c@{\hspace{1.0mm}}c@{\hspace{1.0mm}}c@{\hspace{2.0mm}}c@{\hspace{1.0mm}}c@{\hspace{1.0mm}}c@{\hspace{1.0mm}}c@{\hspace{1.0mm}}c@{\hspace{2.0mm}}c}
\tabletypesize{\scriptsize}
\tablewidth{0pt}
\tablecaption{{\it [Table~2 in \cite{Blondin/etal:2009}].} Reddening and \naoned\ EW measurements\label{tab:ewebv}}
\tablehead{
\colhead{SN} &
\colhead{$E(B-V)$} &
\colhead{EW} &
\colhead{$\Delta{\rm EW}$} &
\colhead{$N_\sigma$} &
\colhead{$\chi^2_\nu$} &
\colhead{dof} &
\colhead{$t_{\rm min}$} &
\colhead{$\Delta t$} \\
 & \colhead{(mag)} & \colhead{(\AA)} & \colhead{(\AA)} & & (const.) &  & (d) & (d) \\
 \colhead{(1)} & \colhead{(2)} &\colhead{(3)} &\colhead{(4)}&\colhead{(5)} &\colhead{(6)} &\colhead{(7)} &\colhead{(8)}&\colhead{(9)}}
\startdata
1994D  & 0.04 (0.01) & 0.11$_{-0.20}^{+0.20}$ & 0.40 (0.24) &  1.6 &  2.10 & 13 &  $-$9.5 &  59.6 \\[.15cm]
1994ae & 0.02 (0.01) & 0.42$_{-0.09}^{+0.16}$ & 0.26 (0.18) &  1.4 &  0.68 & 10 &    +1.4 &  38.6 \\[.15cm]
1995al & 0.06 (0.02) & 1.12$_{-0.11}^{+0.10}$ & 0.21 (0.21) &  1.0 &  0.84 &  8 &    +7.0 &  61.5 \\[.15cm]
1997bp & 0.19 (0.03) & 1.80$_{-0.28}^{+0.26}$ & 0.54 (0.23) &  2.3 &  1.53 &  7 &  $-$1.1 &  28.6 \\[.15cm]
1997br & 0.27 (0.04) & 0.63$_{-0.11}^{+0.05}$ & 0.15 (0.24) &  0.6 &  0.25 &  4 &  $-$7.4 &  28.7 \\[.15cm]
1997do & 0.10 (0.03) & 0.28$_{-0.25}^{+0.12}$ & 0.37 (0.21) &  1.8 &  0.61 &  8 &  $-$6.1 &  27.5 \\[.15cm]
1997dt & 0.61 (0.07) & 1.06$_{-0.13}^{+0.24}$ & 0.37 (0.24) &  1.6 &  1.02 &  3 &  $-$7.9 &   8.9 \\[.15cm]
1998aq & 0.01 (0.01) & 0.23$_{-0.16}^{+0.12}$ & 0.28 (0.18) &  1.5 &  0.60 & 20 &  $-$9.0 &  87.6 \\[.15cm]
1998bu & 0.34 (0.04) & 0.37$_{-0.15}^{+0.13}$ & 0.28 (0.16) &  1.7 &  0.62 & 26 &  $-$2.7 &  59.8 \\[.15cm]
1998dh & 0.17 (0.02) & 0.36$_{-0.11}^{+0.17}$ & 0.28 (0.28) &  1.0 &  0.29 &  4 &  $-$8.8 &   8.8 \\[.15cm]
1998dm & 0.34 (0.05) & 0.60$_{-0.12}^{+0.12}$ & 0.23 (0.27) &  0.9 &  0.22 &  5 &    +5.8 &  18.9 \\[.15cm]
1998es & 0.07 (0.02) & 1.57$_{-0.22}^{+0.15}$ & 0.37 (0.24) &  1.6 &  0.36 & 15 &  $-$6.0 &  36.4 \\[.15cm]
1999cl & 1.20 (0.07) & 3.52$_{-0.26}^{+0.17}$ & 0.43 (0.14) &  3.1 &  2.07 &  9 &  $-$7.5 &  15.8 \\[.15cm]
1999dq & 0.12 (0.03) & 1.19$_{-0.13}^{+0.17}$ & 0.30 (0.22) &  1.4 &  0.45 & 15 &  $-$9.6 &  39.3 \\[.15cm]
2001V  & 0.03 (0.02) & 1.44$_{-0.28}^{+0.26}$ & 0.54 (0.28) &  2.0 &  0.72 & 14 &  $-$9.3 &  33.4 \\[.15cm]
2001en & 0.07 (0.04) & 1.12$_{-0.11}^{+0.16}$ & 0.27 (0.31) &  0.9 &  0.35 &  4 &    +2.7 &   8.7 \\[.15cm]
2001ep & 0.14 (0.04) & 0.74$_{-0.21}^{+0.11}$ & 0.32 (0.27) &  1.2 &  0.36 & 10 &  $-$2.7 &  10.8 \\[.15cm]
2002bo & 0.47 (0.04) & 2.61$_{-0.22}^{+0.27}$ & 0.49 (0.20) &  2.4 &  1.09 & 21 &  $-$7.0 &  36.7 \\[.15cm]
2002cr & 0.09 (0.04) & 0.38$_{-0.15}^{+0.16}$ & 0.30 (0.22) &  1.4 &  0.48 &  4 &  $-$7.4 &  12.8 \\[.15cm]
2002fk & 0.03 (0.02) & 0.12$_{-0.13}^{+0.14}$ & 0.28 (0.23) &  1.2 &  0.87 &  7 &  $-$0.6 &  31.6 \\[.15cm]
2003cg & 1.06 (0.03) & 5.54$_{-0.29}^{+0.27}$ & 0.56 (0.26) &  2.2 &  0.97 & 11 &  $-$5.6 &  15.9 \\[.15cm]
2003du & 0.01 (0.01) & 0.10$_{-0.22}^{+0.17}$ & 0.39 (0.22) &  1.8 &  0.77 & 12 &  $-$9.7 &  38.8 \\[.15cm]
2003kf & 0.04 (0.03) & 0.52$_{-0.25}^{+0.16}$ & 0.41 (0.20) &  2.0 &  0.38 & 14 &  $-$7.3 &  48.4 \\[.15cm]
2005am & 0.03 (0.02) & 0.16$_{-0.22}^{+0.14}$ & 0.37 (0.21) &  1.7 &  0.60 &  7 &    +0.4 &   8.9 \\[.15cm]
2005cf & 0.10 (0.04) & 0.17$_{-0.18}^{+0.35}$ & 0.53 (0.28) &  1.9 &  1.03 & 23 & $-$11.8 &  41.7 \\[.15cm]
2006N  & 0.03 (0.02) & 0.68$_{-0.31}^{+0.33}$ & 0.65 (0.28) &  2.3 &  1.21 &  7 &  $-$2.8 &  11.8 \\[.15cm]
2006X  & 1.47 (0.04) & 1.92$_{-0.41}^{+0.36}$ & 0.78 (0.17) &  4.5 &  3.68 & 18 &    +1.7 &  58.5 \\[.15cm]
2007S  & 0.41 (0.03) & 1.92$_{-0.25}^{+0.34}$ & 0.59 (0.25) &  2.3 &  1.20 &  6 &  $-$4.7 &  29.6 \\[.15cm]
2007af & 0.14 (0.03) & 0.45$_{-0.19}^{+0.21}$ & 0.40 (0.20) &  1.9 &  0.86 & 24 &  $-$4.7 &  97.2 \\[.15cm]
2007bm & 0.52 (0.04) & 2.28$_{-0.26}^{+0.14}$ & 0.40 (0.16) &  2.5 &  1.34 &  6 &  $-$9.3 &  29.8 \\[.15cm]
2007ca & 0.30 (0.04) & 1.92$_{-0.10}^{+0.12}$ & 0.22 (0.31) &  0.7 &  0.16 &  3 &    +3.4 &   9.9 
\enddata
\tablecomments{Col.~(1): SN name; col.~(2): host-galaxy color excess determined
from fits to multi-band optical light curves using the MLCS2k2 code
of \cite{MLCS2k2}; col.~(3): weighted mean EW. The upper and lower
limits correspond to the maximum deviations from the weighted mean;
col.~(4): maximum EW difference. The $1\sigma$ error appears in
between parentheses; col.~(5): $\Delta{\rm EW}$ divided by its
$1\sigma$ error; col.~(6): $\chi^2$ per degree of freedom for a
constant EW fit; col.~(7): number of degrees of freedom (simply the
number of data points minus one); col.~(8): age (in days from $B$-band
maximum light) of the earliest spectrum; col.~(9): age range (in days)
of the spectra used in the fit.}
\end{deluxetable}

\bibliographystyle{apj}
\bibliography{ms_erratum}